\documentclass[showpacs,aps,prd,nofootinbib,showkeys,superscriptaddress,twocolumn,floatfix]{revtex4}
\usepackage{graphicx}
\usepackage{bm}
\usepackage{amssymb}
\usepackage{amsmath,latexsym}
\usepackage[usenames]{color}
\usepackage{subfigure}
\usepackage{subfigure}
\usepackage{physymb}
\usepackage{slashed}
\usepackage{multirow,array}
\usepackage{mathtools}
\usepackage{mathrsfs}
\usepackage[colorlinks=false,linktocpage=true]{hyperref}

\pdfoutput=1
\newcommand{\be}{\begin{equation}}
\newcommand{\ee}{\end{equation}}
\newcommand{\bea}{\begin{eqnarray}}
\newcommand{\eea}{\end{eqnarray}}

\newcommand{\st}{\sigma_T}

\newcommand{\bin}{\binom {n+1}{r}}

\newcommand{\eq}{{\,=\,}}

\begin{document}

\title{Nonlinear dynamics from the relativistic Boltzmann equation in the
Friedmann-Lema\^itre-Robertson-Walker spacetime}
\date{\today}
\author{D.~Bazow}
\affiliation{Department of Physics, The Ohio State University, Columbus, OH 43210, USA}
\author{G.~S.~Denicol}
\affiliation{Instituto de F\'isica, Universidade Federal Fluminense, UFF, Niter\'oi, 24210-346, RJ, Brazil}
\affiliation{Department of Physics, Brookhaven National Laboratory, Upton, New York
11973-5000}
\author{U.~Heinz}
\affiliation{Department of Physics, The Ohio State University, Columbus, OH 43210, USA}
\author{M.~Martinez}
\affiliation{Department of Physics, The Ohio State University, Columbus, OH 43210, USA}
\author{J.~Noronha}
\affiliation{Instituto de F\'{\i}sica, Universidade de S\~{a}o Paulo, C.P. 66318,
05315-970 S\~{a}o Paulo, SP, Brazil}

\begin{abstract}
The dissipative dynamics of an expanding massless gas with constant cross section in a spatially flat Friedmann-Lema\^itre-Robertson-Walker (FLRW) universe is studied. The mathematical problem of solving the full nonlinear relativistic Boltzmann equation is recast into an infinite set of nonlinear ordinary differential equations for the moments of the one-particle distribution function. Momentum-space resolution is determined by the number of non-hydrodynamic modes included in the moment hierarchy, i.e., by the truncation order. We show that in the FLRW spacetime the non-hydrodynamic modes decouple completely from the hydrodynamic degrees of freedom. This results in the system flowing as an ideal fluid while at the same time producing entropy. The solutions to the nonlinear Boltzmann equation exhibit transient tails of the distribution function with nontrivial momentum dependence. The evolution of this tail is not correctly captured by the relaxation time approximation nor by the linearized Boltzmann equation. However, the latter probes additional high-momentum details unresolved by the relaxation time approximation. While the expansion of the FLRW spacetime is slow enough for the system to move towards (and not away from) local thermal equilibrium, it is not sufficiently slow for the system to actually ever reach complete local equilibrium. Equilibration is fastest in the relaxation time approximation, followed, in turn, by kinetic evolution with a linearized and a fully nonlinear Boltzmann collision term.
\end{abstract}

\maketitle

\section{Introduction}
\label{sec:intr} 

The Boltzmann equation is the main theoretical framework for studying the dissipative out-of-equilibrium dynamics of dilute gases. Within this approach, the transport and thermodynamic properties of matter are understood in terms of the one-particle distribution function whose phase-space evolution is determined by the Boltzmann equation. The physics and the mathematics involved in the \emph{non-relativistic} Boltzmann equation have been thoroughly studied~\cite{Cercignani-1, Cercignani-2} and, in certain limits, analytical solutions of this nonlinear integro-differential equation are known. 

For instance, Bobylev~\cite{bobylev}, Krook, and Wu (BKW)~\cite{KW-1,KW-2} derived an exact solution of the Boltzmann equation that describes the nonlinear relaxation of a non-expanding, non-relativistic homogeneous gas with elastic cross section inversely proportional to the relative speed. For this case it was shown that a generic solution to the Boltzmann equation can be obtained in terms of the moments of the distribution function whose temporal evolution is dictated by a coupled set of nonlinear ordinary differential equations. A remarkable feature of the BKW solution is the formation of transient high energy tails due to the nonlinear mode-by-mode coupling among different moments of the distribution function. These high energy tails show how high energy moments of the distribution are populated over time, a process that directly affects the relaxation of the distribution function towards global equilibrium.

The \emph{relativistic} generalization of the Boltzmann equation is an active topic of research that has applications in different areas of physics, ranging from thermal field theory~\cite{Jeon:1995zm, Blaizot:2001nr, kapusta, lebellac, Litim:2001db, Litim:1999id, Blaizot:1999xk} to high-energy nuclear collisions \cite{Mathieu:2014aba, Jeon:2004dh, Epelbaum:2014mfa, Heinz:1985qe, Heinz:1983nx, Elze:1989un, Mueller:2002gd, Gelis:2007pw, Bass:1998ca, Molnar:2001ux, Xu:2004mz, Epelbaum:2015vxa}, cosmology \cite{Weinberg-1, Weinberg-2, Cercignani, Bernstein, DeGroot:1980dk, Dodelson} and astrophysics \cite{Ma:1995ey, Liebendoerfer:2003es, Janka:2006fh, Oldengott:2014qra}. A major topic of interest in relativistic kinetic theory is to quantify the role of nonlinear effects in rapidly expanding plasmas, which requires a careful analysis of the type of interactions between the constituent particles of the system that defines the collision kernel. In practice, this kinetic equation is solved numerically, although it is possible to find exact solutions of the relativistic Boltzmann equation for highly symmetric systems using the relaxation time approximation as a model for the collision term \cite{Cercignani, Baym:1984np, Denicol:2014tha, Denicol:2014xca, Noronha:2015jia, Hatta:2015kia, Nopoush:2014qba} that describes the relaxation of the system to its equilibrium state with a single microscopic time scale. These exact solutions have been extremely useful to understand certain features of the thermalization process in relativistic gases while also providing nontrivial ways to test the accuracy and precision of numerical algorithms for solving the Boltzmann equation and macroscopic hydrodynamic approximations to the microscopic kinetic evolution\cite{Welke:1989dr, Florkowski:2013lya, Marrochio:2013wla, Denicol:2014tha, Denicol:2014xca,
Nopoush:2014qba}. 

However, a complete understanding of the
dissipative dynamics of an expanding gas can only be achieved by solving the
full nonlinear Boltzmann equation which necessarily embodies an entire hierarchy of
microscopic relaxation time scales and includes mode-by-mode coupling effects. For instance, 
it has been shown that nonlinear effects play an important role in the
hydrodynamization process of the quark gluon plasma at weak coupling~\cite{Berges:2008wm,
Dusling:2010rm, Denicol:2012cn, Berges:2013eia, Berges:2013fga,
Berges:2013lsa, Gelis:2013rba, Molnar:2013lta, Kurkela:2014tea,
Berges:2014bba, Kurkela:2015qoa, Berges:2015ixa, Epelbaum:2015vxa} and in the
reheating process of inflationary cosmology~\cite{Polarski:1995jg,
Khlebnikov:1996mc, Prokopec:1996rr, Son:1996uv, Son:1996zs, Micha:2002ey,
Micha:2004bv, Frolov:2008hy, Allahverdi:2010xz}. While these effects have been studied only numerically, it would be extremely useful to also investigate the nonlinear out-of-equilibrium dynamics of rapidly expanding systems analytically where this is possible.

The first step towards this goal was taken in \cite{Bazow:2015dha} by recasting the general relativistic Boltzmann equation, in a spatially homogeneous and isotropically expanding Friedmann-Lema\^itre-Robertson-Walker (FLRW) universe, in terms of ordinary nonlinear differential equations for the energy moments of the distribution function. There these moment equations were solved analytically for a very specific far-from-equilibrium initial condition, and the corresponding distribution function was found. This led to a new class of analytical solutions of the relativistic Boltzmann equation. Moreover, as observed in \cite{Bazow:2015dha}, the symmetries of the FLRW spacetime restrict the energy momentum tensor to ideal fluid form, whether or not the system is in local thermal equilibrium. The macroscopic hydrodynamic quantities (energy density, temperature, and hydrodynamic flow) thus evolve according to the laws of ideal fluid dynamics while the system, if initialized in a non-equilibrium state, produces entropy. This provides an explicit counter example to the folklore that a system must be in local thermal equilibrium for the hydrodynamic currents to exhibit ideal fluid behavior. 

In this paper we obtain semi-analytical (numerical) solutions to the moment equations in FLRW spacetime for various initial conditions, to arbitrary accuracy. These new solutions are used to investigate the domain of applicability of two widely used approximation schemes for the Boltzmann equation: the linearized Boltzmann collision term and the relaxation time approximation. Such a study not only gives insight into the physical features neglected in these two approximations but it also illustrates how mode-by-mode nonlinear coupling dynamics manifests itself within the relativistic Boltzmann equation.

This article is organized as follows: In Sec.~\ref{sec:FLRWspace} we briefly review some of the basic properties of the FLRW metric and introduce our notation. In the rest of Sec.~\ref{sec:FLRWmethod} we provide a detailed derivation of the general method that allows one to find exact solutions to the nonlinear Boltzmann equation in the FLRW spacetime. We refer to our previous result~\cite{Bazow:2015dha} in Sec.~\ref{sec:exact} where we re-derive an exact solution to the Boltzmann equation valid for a specific far-from-equilibrium initial condition. We discuss the entropy production of this system in Sec.~\ref{sec:entprod}. Results from numerical studies involving the different evolution schemes for the distribution function and the mode-by-mode coupling effects are shown in Sec.~\ref{sec:numres}. A summary of our findings and some general conclusions are presented in Sec.~\ref{sec:concl}. Some technical details of the calculations can be found in the appendices.

\section{Exact solution of the Boltzmann equation in an FLRW spacetime}
\label{sec:FLRWmethod} 

The existence and uniqueness of a solution to the relativistic Boltzmann equation in the spatially flat FLRW spacetime has been demonstrated in Refs.~\cite{Lee20134267,Lee:2013ig}. However, until Ref.\cite{Bazow:2015dha}, no explicit analytical solution was known. Building on the work performed in Ref.\cite{Bazow:2015dha}, we here continue to study nonlinear effects in the Boltzmann equation for an expanding gas of massless particles. Our starting point is the Boltzmann equation for a relativistic massless gas that expands isotropically and homogeneously in a FLRW spacetime. We replace the Boltzmann equation by an infinite hierarchy of equations for its moments -- a set of coupled ordinary differential equations for moments of the distribution function. This hierarchy can then be solved, for any initial condition and to an arbitrary precision, by truncation at an appropriate order. Finally, the distribution function may be reconstructed from the moments.

\subsection{Notation and some properties of FLRW spacetime}
\label{sec:FLRWspace} 

The FLRW metric is a solution to Einstein's equations describing a spatially
homogeneous and isotropically expanding universe \cite{Weinberg-1,
Weinberg-2, Bernstein, baumann}. For a spatially flat universe the FLRW
metric reads 
\begin{equation}
ds^{2}=dt^{2}-a^{2}(t)\,\gamma _{ij}\,dx^{i}\,dx^{j}  \label{eq:FLRWds}
\end{equation}
where $i,j\in \{1,2,3\}$, $\gamma _{ij}$ is the spatial metric of the 3-dimensional space, and $a(t)$ is a dimensionless scale factor accounting for the expansion that is determined by solving Einstein's equations. The general form of $\gamma _{ij}$ depends on the choice of coordinates; in this work, we use spatial Cartesian coordinates, $\gamma_{ij}=\delta _{ij}$.\footnote{%
     The FLRW metric~\eqref{eq:FLRWds} is not the most general metric for a maximally 
     symmetric space. For instance, in a spatially curved 3-space with constant Gaussian 
     curvature $K$ the line element of FLRW spacetime is given in polar coordinates by 
     \begin{equation*}
     ds^{2}=dt^{2}-a^{2}(t)\left[ \frac{dr^{2}}{1-K\,r^{2}}+r^{2}\,d\Omega ^{2}\right].
     \end{equation*}
}
The determinant of this metric is $g\equiv\mathrm{det}(g_{\mu\nu})=-a^6(t)$ such that $\sqrt{-g}\eq{a}^3(t)$.

The FLRW metric~\eqref{eq:FLRWds} is invariant under the following
transformation: 
\begin{equation}
   x^{i}\rightarrow x^{i}/\lambda \,,\hspace{1cm}a(t)\rightarrow \lambda\,a(t)\,  
   \label{eq:rescFLRW}
\end{equation}
and, due to this scaling symmetry, one can set $a(t_{0})=1$ at the initial time $t_{0}$ (which we choose as $t_0\eq0$). This will be our boundary condition for the scaling factor.

We denote the scalar product between 4-vectors, $a^\mu$, $b^\mu$, as $a{\,\cdot\,}b \equiv a_\mu b^\mu$. For massless particles with 4-momentum $k^{\mu }$, the on-shell condition, $k{\,\cdot\,}k\eq0$, yields $k^{0}\eq{a(t)}\sqrt{(k^1)^2{+}(k^2)^2{+}(k^3)^2}$ where $(k^1,k^2,k^3)\equiv(k_x,k_y,k_z)$ is the usual
3-momentum, given by the spatial contravariant components of the 4-vector $k^\mu$. Following \cite{Debbasch1:2009, Debbasch2:2009} we find it convenient to express this instead in terms of the magnitude $k$ of a 3-vector ${\bf k}\equiv(k_1,k_2,k_3)$ constructed from the {\em covariant} spatial components $k_i$ of the 4-momentum, $k\eq\sqrt{k_1^2{+}k_2^2{+}k_3^2}$. In terms of $k$ the on-shell condition for massless particles reads
\be
\label{eq3}
   k^0{\eq}k/a(t).
\ee
This way of expressing $k^{0}$ is rather convenient since, as we shall see in the following section, the factor $a(t)$ in Eq.~(\ref{eq3}) will cancel in the exponent of the equilibrium Boltzmann distribution function.

Furthermore, the Lorentz-invariant momentum space integration measure in curved spacetime is \cite{Debbasch1:2009, Debbasch2:2009}
\begin{equation}
\sqrt{-g}\,d^{4}k\equiv \sqrt{-g}\,dk^{0}dk^{1}dk^{2}dk^{3}=\frac{%
dk_{0}dk_{1}dk_{2}dk_{3}}{\sqrt{-g}},  \label{eq4}
\end{equation}
while the Lorentz-covariant 3-momentum integration measure over on-shell distributions can be written as
\begin{eqnarray} 
\label{eq5}
   \int_k &\equiv& \frac{\sqrt{-g}} {(2\pi)^3} 
                            \int 2 \theta(k^0) \,\delta(k{\cdot}k{-}m^2)\,dk^0 dk^1 dk^2 dk^3 
\nonumber\\
   &=& \int \frac{dk_1 dk_2 dk_3} {(2\pi)^3k^0\sqrt{-g}}
           \equiv \frac{1}{\bigl(2\pi a(t)\bigr)^3}\int \frac{d^3k}{k^0},
\end{eqnarray}
where the last equality defines our notation $d^3k\equiv dk_1 dk_2 dk_3$ in terms of the {\em covariant} spatial components of the momentum four-vector. Using spherical coordinates, this reduces for massless particles to
\be
\label{eq6}
   \int_k = \frac{1}{(2\pi)^3} \int_0^\infty \frac{k\,dk}{a^2(t)} \int d\Omega_k,
\ee
with $k$ as defined above. 

\subsection{The relativistic Boltzmann equation in FLRW spacetime}
\label{subsec:Boltzeq-FLRW} 

The general relativistic Boltzmann equation for an on-shell single-particle distribution function $f(x,k)$ is given by~\cite{Debbasch1:2009, Debbasch2:2009, Cercignani, DeGroot:1980dk} 
\begin{equation}
k^{\mu }\bigl(u_{\mu }D{+}\nabla _{\mu }\bigr)f(x,k
)+k_{\lambda }k^{\mu }\Gamma _{\mu i}^{\lambda }\frac{\partial f(x,k)}{\partial k_{i}}=\mathcal{C}[f],  \label{eq:Boltzmann1}
\end{equation}
where $C[f]$\ is the nonlinear collision term for binary collisions, and $\Gamma _{\mu \nu }^{\lambda }=\frac{1}{2}g^{\lambda \gamma }\left( \partial _{\mu}g_{\gamma \nu }{+}\partial _{\nu }g_{\gamma \mu }{-}\partial _{\gamma}g_{\mu \nu }\right)$\ are the Christoffel symbols. In \eqref{eq:Boltzmann1} we have decomposed the space-time derivative $\partial_{\mu}$ into its temporal and spatial components in the comoving frame, $\partial_{\mu}{\,=\,}u_{\mu }D{+}\nabla_{\mu}$, with $D{\,\equiv\,}u_{\nu}\partial^{\nu}$, $\nabla_{\mu}{\,\equiv\,}\Delta_{\mu\nu}\partial^{\nu}$. Here, $u^{\mu }=(1,0,0,0)$ is the 4-velocity of the comoving frame and $\Delta _{\mu \nu }\,\equiv \,\,g_{\mu \nu }-u_{\mu }u_{\nu }$\ the
projection operator onto the spatial components in this frame.%
\footnote{%
      Note that even though a fluid filling a FLRW universe homogeneously is locally static, the expanding 
      FLRW geometry induces a nonzero fluid expansion rate $\theta (t)\equiv \partial _{\mu }
      (\sqrt{-g}\,u^{\mu })/\sqrt{-g}=3H(t)$ where $g=-a^{6}(t)$ is the determinant of the FLRW metric 
      and $H(t)=\dot{a}(t)/a(t)$ is the Hubble parameter.
      }

The symmetries of the system restrict the number of independent variables
upon which the distribution function can depend~\cite{Baym:1984np,
Denicol:2014xca, Denicol:2014tha, Noronha:2015jia, Hatta:2015kia}. In our
case, the homogeneity of the FLRW spacetime \eqref{eq:FLRWds} implies that,
in the comoving frame, the distribution function $f(x,k)\to f(t,k)$ is independent of the spatial coordinates and
is spherically symmetric in momentum space \cite{Bernstein, Weinberg:1971mx}. For a general collision kernel, we define the shorthand notation $f_k(t)\equiv f(t,k)$ for the distribution function and the Boltzmann equation in FLRW spacetime
thus reads\footnote{%
   In the comoving frame the only non-zero Christoffel symbols are 
   $\Gamma _{ij}^{0}=a(t)\dot{a}(t)\,\delta _{ij}$ and $\Gamma _{0j}^{i}=\delta _{j}^{i}\,H(t)$. 
   For the FLRW spacetime, the term in \eqref{eq:Boltzmann1} involving the Christoffel 
   symbols thus cancels exactly: 
\begin{equation*}
k_{\lambda }k^{\mu }\Gamma _{\mu i}^{\lambda }\frac{\partial f_k}{%
\partial k_{i}}=\left( \Gamma _{ji}^{0}+g_{lj}\Gamma _{0i}^{l}\right)
k^{0}\,k^{j}\frac{\partial f_k}{\partial k_{i}}=0\,.
\end{equation*}%
}
\begin{equation}
(u\cdot k)\,Df_k=\mathcal{C}[f]\,,  \label{eq:FLRWBoltzgen}
\end{equation}
where, in the comoving frame, $u\cdot k=k^{0}{\,=\,}k/a(t)$.

For a single particle species with classical Boltzmann statistics, the
collision term $\mathcal{C}[f]$ takes the form \cite{DeGroot:1980dk} 
\begin{equation}
 \mathcal{C}[f]=\frac{1}{2}\int_{k^{\prime }pp^{\prime }}\,W_{\mathbf{k} 
 \mathbf{k^{\prime }}\rightarrow \mathbf{p}\mathbf{p^{\prime }}}\,
 \left( f_{p}f_{p^{\prime }}-f_kf_{k^{\prime }}\right) ,  
 \label{eq:collision1}
\end{equation}
where $W_{\mathbf{k}\mathbf{k^{\prime }}\rightarrow \mathbf{p}\mathbf{%
p^{\prime }}}$ is the transition rate and $\int_{p}$ is defined as in Eq.~(%
\ref{eq5}) in terms of the covariant spatial components of the
momentum $p$ in the comoving frame. The transition rate can be written in
terms of the differential cross section $\sigma (s,\Theta )$ as follows \cite%
{DeGroot:1980dk, Debbasch1:2009, Debbasch2:2009}: 
\begin{equation}
W_{\mathbf{k}\mathbf{k^{\prime }}\rightarrow \mathbf{p}\mathbf{p^{\prime }}%
}=s\,\sigma (s,\Theta _{s})\,(2\pi )^{6}\sqrt{-g}\,\delta ^{4}(k{+}k^{\prime
}{-}p{-}p^{\prime }).  \label{eq:transition}
\end{equation}
Here the total energy $s$ and the scattering angle $\Theta _{s}$ are given by
\begin{equation}
  s=(k {+}k^{\prime })\cdot (k {+} k^{\prime}),\quad \cos\Theta _{s}
    =\frac{(k {-}k^{\prime })\cdot (p {-}p^{\prime})}{(k {-}k^{\prime })\cdot (k{-}k^{\prime})}\,.
\end{equation}
The transition rate $W_{\mathbf{k}\mathbf{k^{\prime }}\rightarrow \mathbf{p}\mathbf{p^{\prime }}}$ in \eqref{eq:transition} is a Lorentz scalar and obeys the detailed balance and crossing symmetries $W_{\mathbf{k}\mathbf{k^{\prime }}\rightarrow \mathbf{p}\mathbf{p^{\prime }}}=W_{\mathbf{p}\mathbf{p^{\prime }}\rightarrow \mathbf{k}\mathbf{k^{\prime }}}{\,=\,}{W}_{\mathbf{k}\mathbf{k}^{\prime }\rightarrow \mathbf{p}^{\prime }\mathbf{p}}$ \cite{DeGroot:1980dk, Peskin:1995ev}.

For simplicity we here assume isotropic scattering, i.e., the differential cross section depends only on $s$. Then we can express the transition rate through the total cross section $\sigma _{T}(s){\,\equiv\,}\pi \int d\Theta _{s}\,\sin\Theta _{s}\,\sigma (s,\Theta _{s})$,\footnote{%
    Note that, due to the indistinguishability of the two particles, we integrate here only over
    half the solid angle, i.e. over $2\pi$.
    }
and the Boltzmann equation in the FLRW spacetime \eqref{eq:FLRWBoltzgen} can be written as 
\begin{equation}
(u\cdot k)\,Df_k=\mathcal{C}_{\mathrm{gain}}-\mathcal{C}_{%
\mathrm{loss}},  \label{eq:Boltzmanneq2}
\end{equation}
with the gain and loss terms 
\begin{subequations}
\label{eq:gain-loss}
\begin{align}
\!\!\!\!\mathcal{C}_{\mathrm{gain}}& =\frac{(2\pi )^{5}}{2}%
\!\!\int_{k^{\prime }pp^{\prime }}\!\!\!\!\!\!\!s\,\sigma _{T}(s)\sqrt{{-}g}%
\,\delta ^{4}(k{+}k^{\prime }{-}p{-}p^{\prime })\,f_{p}f_{p^{\prime}},\!\!
\label{eq:loss}\\
\!\!\!\!\mathcal{C}_{\mathrm{loss}}& =\frac{(2\pi )^{5}}{2}%
\!\!\int_{k^{\prime }pp^{\prime }}\!\!\!\!\!\!\!s\,\sigma _{T}(s)\sqrt{-g}%
\,\delta ^{4}(k{+}k^{\prime }{-}p{-}p^{\prime })\,f_kf_{k^{\prime }}.\!\!
\end{align}
\end{subequations}
In the following subsections we replace all of the physical information contained in the Boltzmann equation (a nonlinear integro-differential equation for the distribution function $f_k$) by a set of equations for the energy moments of the distribution function.


\subsection{Normalized energy moments and their evolution equations}


We define the energy moments $\rho _{n}$ of the distribution function as follows%
\footnote{%
     In kinetic theory, it is usually assumed that the distribution function $f_{k}$ belongs to the 
     Hilbert space $L_{2}(0,\infty )$, i.e., the space of square-integrable functions defined in 
     the interval $k\in (0,\infty)$ \cite{DeGroot:1980dk, Ernst-2, bobylev}. In this case, it is then 
     guaranteed that the moments $\rho _{n}$~\eqref{eq:rhomoments} are finite.
     } 
\begin{equation}
\rho _{n}(t)=\int_{k}(u\cdot k)^{n+1}\,f_k=\frac{1}{2\pi
^{2}}\int_{0}^{\infty }\frac{dk\,k^{n+2}}{a^{n+3}(t)}\,f_k.
\label{eq:rhomoments}
\end{equation}
The positivity of the distribution function implies that $\rho _{n}(t)\geq 0$. The number and energy densities are given by $\rho _{0}(t)$ and $\rho_{1}(t)$, respectively. The higher-order moments $\rho _{n\geq 2}(t)$ do not have an intuitive macroscopic interpretation but are needed to resolve additional microscopic details of the system. Moments of lower order $n$ correspond to softer momentum modes (longer wavelengths) while moments of higher-order probe the short wavelength structure of the local distribution function.

The collision kernel in \eqref{eq:collision1} conserves particle number, energy, and momentum. In an FLRW spacetime the corresponding moments $\rho_{0}(t)$ and $\rho _{1}(t)$ evolve by the following equations \cite{baumann}:
\begin{subequations}
\label{eq:conservation}
\begin{align}
& D\rho _{0}(t)+3\rho _{0}(t)H(t)=0\,,  \label{eq:conserv-ener} \\
& D\rho _{1}(t)+4\rho _{1}(t)H(t)=0\,.
\end{align}
\end{subequations}
Equations~(\ref{eq:conservation}) follow from Einstein's equations for a homogeneous and isotropic fluid in an FLRW metric. They correspond to the equations of motion of an ideal fluid, $d_\mu j_{\mu }^{\mu }{\,=\,}d_\mu T^{\mu \nu }{\,=\,}0$ (where $d_\mu$ denotes the covariant derivative), with particle current $j^{\mu }(t)\eq{n(t)}u^{\mu }$ and energy-momentum tensor $T^{\mu \nu }(t)=e(t)\,\bigl(\frac{4}{3}u^{\mu}u^{\nu }{-}\frac{1}{3}g^{\mu \nu }\bigr)$. Equations (\ref{eq:conservation}) are solved by
\begin{subequations}
\label{eq:sol-conservation}
\begin{align}
\rho _{0}(t)\,& =\frac{n_{0}}{a^{3}(t)}=\frac{1}{a^{3}(t)}\frac{\lambda
_{0}T_{0}^{3}}{\pi ^{2}}\,, \\
\rho _{1}(t)\,& =\frac{e_{0}}{a^{4}(t)}=\frac{1}{a^{4}(t)}\frac{3\lambda
_{0}T_{0}^{4}}{\pi ^{2}}\,,
\end{align}
\end{subequations}
where $n_{0}\equiv \rho _{0}(0)$ and $e_{0}\equiv \rho _{1}(0)$ are the initial particle and energy densities, and $T_{0}$ and $\lambda _{0}$ are the initial temperature and fugacity assigned to the system. The temperature and fugacity of our nonequilibrium system are obtained from the matching conditions 
\begin{subequations}
\label{eq:sol-conservationEQ}
\begin{align}
\rho _{0}(t)\,& =n^{\mathrm{eq}}(t)=\,\frac{\lambda (t)T^{3}(t)}{\pi ^{2}}\,,
\label{rho1} \\
\rho _{1}(t)\,& =e^{\mathrm{eq}}(t)=\frac{3\lambda (t)T^{4}(t)}{\pi ^{2}}\,.
\end{align}
\end{subequations}
By comparing Eqs.~(\ref{eq:sol-conservation}) and (\ref{eq:sol-conservationEQ}) we find $\lambda {\,=\,}\mathrm{constant}$ and $T(t){\,=\,}{T}_{0}/a(t)$, such that the local equilibrium distribution function has the following form (remember that $u\cdot k=k/a(t)$ in the comoving frame)
\begin{equation}
f_k^{\mathrm{eq}}=\lambda (t)\,e^{-u\cdot k/T(t)}=\lambda
\,e^{-k/T_{0}}.  \label{eq:distreq-1}
\end{equation}
One sees that, when $f_k^{\mathrm{eq}}$ is expressed in terms of the magnitude $k$ of the covariant spatial components of the momentum four-vector, its dependence on $a(t)$ completely cancels (hence\  $f_k^{\mathrm{eq}}$ is time independent).\footnote{%
      The physics of this is the following \cite{baumann,Weinberg-2}: A comoving observer defines 
      the {\em physical} 3-momentum of a massless particle via the energy-momentum relation 
      $k^0=E_\mathrm{phys}=|{\bf k}_\mathrm{phys}|$. The discussion in Sec.~\ref{sec:FLRWspace} 
      shows that this physical 3-momentum ${\bf k}_\mathrm{phys}$ is related to the {\em covariant} 
      spatial components of the momentum 4-vector by ${\bf k}_\mathrm{phys}={\bf k}/a(t)$, and its 
      magnitude $k_\mathrm{phys}$ is related to the magnitude $k$ of the {\em covariant} components 
      of the momentum four-vector by $k_\mathrm{phys} = k/a(t)$. Hence 
      $k/T_0 = k_\mathrm{phys}/T(t) = E_\mathrm{phys}/T(t)$ where 
      $T(t)=T_0/a(t)$ is the cosmologically redshifted temperature of the expanding FLRW universe 
      as seen by the comoving observer.
      }
For later convenience we also introduce the energy moments of the equilibrium distribution function: 
\begin{equation}
\rho _{n}^{\mathrm{eq}}(t)\equiv \int_{k}\,(u\cdot k)^{n+1}\,f_k^{\mathrm{eq}}(t)=\frac{(n{+}2)!}{2\pi ^{2}}\,\lambda \,T^{n+3}(t)\,.
\label{eq:rhomomeq}
\end{equation}

We now use the Boltzmann equation to derive the set of equations of motion
satisfied by the energy moments $\rho _{n}(t)$. To this end we apply the
comoving time derivative $D$ to the definition of $\rho _{n}$ and substitute
the resulting time derivative of the distribution function $Df_k$
from Eq.~\eqref{eq:Boltzmanneq2}. This results in the following evolution
equation for the moments $\rho _{n}$: 
\begin{equation}
D\rho _{n}(t)+(3+n)H(t)\rho _{n}(t)=\mathcal{C}_{\mathrm{gain}}^{(n)}(t)-%
\mathcal{C}_{\mathrm{loss}}^{(n)}(t)\,,  \label{eq:rhoequation}
\end{equation}%
where the $n^{\mathrm{th}}$ moments of the loss and gain terms, $\mathcal{C}%
_{\mathrm{loss}}^{(n)}$ and $\mathcal{C}_{\mathrm{gain}}^{(n)}$,
respectively, are given by the following expressions: 
\begin{subequations}
\label{eq:coll-mom}
\begin{align}
\mathcal{C}_{\mathrm{loss}}^{(n)}=& \frac{(2\pi )^{5}}{2}\int_{kk^{\prime
}pp^{\prime }}s\,\sigma _{T}(s)\,(u{\,\cdot\,}k)^{n}\,  
\notag \label{eq:loss1} \\
& \times \sqrt{-g}\,\delta ^{4}(k{+}k^{\prime }{-}p{-}p^{\prime })\,f_k f_{k'}\,, \\
\label{eq:gain1}
\mathcal{C}_{\mathrm{gain}}^{(n)}=& \frac{(2\pi )^{5}}{2}\,\int_{kk^{\prime
}pp^{\prime }}s\,\sigma _{T}(s)\,(u{\,\cdot\,}p)^{n}\,  
\notag \\
& \times \,\sqrt{-g}\,\delta ^{4}(k{+}k^{\prime }{-}p{-}p^{\prime })\,f_k f_{k'}\,.
\end{align}%
For an energy independent total cross section $\sigma _{T}(s)=\mathrm{const}$
(\textquotedblleft hard sphere approximation\textquotedblright ) the
integrals in Eq.~\eqref{eq:coll-mom} can be done analytically (see Appendix %
\ref{app:mom-collker}), with the result 
\end{subequations}
\begin{subequations}
\label{eq:coll-mom2}
\begin{align}
\mathcal{C}_{\mathrm{loss}}^{(n)}(t)& =\sigma _{T}\rho _{n}(t)\rho _{0}(t)\,,
\label{eq:gain2} \\
\mathcal{C}_{\mathrm{gain}}^{(n)}(t)& =2\sigma _{T}\sum_{m=0}^{n}\frac{(n{+}%
2)\,n!}{(m{+}2)!(n{-}m{+}2)!}\,\rho _{n-m}(t)\rho _{m}(t)\,.
\end{align}
\end{subequations}
Substituting these results in Eq.~\eqref{eq:rhoequation} one obtains the
following set of coupled evolution equations for the moments $\rho _{n}$,
which is equivalent to the Boltzmann equation: 
\begin{equation}
\begin{split}
& D\rho _{n}(t)+(3{+}n)H(t)\rho _{n}(t)+\sigma _{T}\rho _{0}(t)\rho _{n}(t)
\\
& =2\sigma _{T}\sum_{m=0}^{n}\frac{(n{+}2)\,n!}{(m{+}2)!(n{-}m{+}2)!}\,\rho
_{n-m}(t)\rho _{m}(t)\,.
\end{split}
\label{eq:rhomomfinal}
\end{equation}
The conservation laws~\eqref{eq:conservation} are recovered by setting $n\eq0$ and $n\eq1$, respectively.

Defining the normalized moments 
\begin{equation}
M_{n}(t)=\frac{\rho _{n}(t)}{\rho _{n}^{\mathrm{eq}}(t)},  \label{eq:defMmom}
\end{equation}%
and substituting them into Eq.~\eqref{eq:rhomomfinal} one obtains a similar
infinite nonlinear hierarchy of coupled ordinary differential equations for
the $M_{n}$ moments: 
\begin{equation}
a^{3}(\hat{t}\,)\,\partial_{\hat{t}}M_{n}(\hat{t}\,)+M_{n}(\hat{t}\,)=\frac{1}{n{+}1%
}\sum_{m=0}^{n}M_{m}(\hat{t}\,)M_{n-m}(\hat{t}\,).  \label{eq:Mmoment}
\end{equation}%
Here we defined the dimensionless time variable $\hat{t}=t/\ell _{0}$ where $%
\ell _{0}=1/(\sigma _{T}n_{0})$ is the mean free path at $t{\,=\,}0$. 

The solution of this infinite set of nonlinear coupled differential
equations~\eqref{eq:Mmoment} contains the same physical information as the
original Boltzmann equation. At the level of moments, the nonlinear
dependence of the collision kernel on the distribution function is encoded
in the mode-by-mode coupling between moments of different order, as seen on
the r.h.s. of Eq.~\eqref{eq:Mmoment}. The conservation laws~%
\eqref{eq:conservation} together with the matching conditions imply that the
only non-evolving moments are $M_{0}(\hat{t})$ and $M_{1}(\hat{t})$: 
\begin{equation}
M_{0}(\hat{t})=M_{1}(\hat{t})=1\quad \mbox{for all }\hat{t}.  \label{M01}
\end{equation}

It is convenient to further express the time dependence of the moments in terms of the variable 
\begin{equation}
\tau \left( \hat{t}\right) =\int_{\hat{t}_{0}}^{\hat{t}}\frac{d\hat{t}%
^{\prime }}{a^{3}(\hat{t}^{\prime })}
\label{eq:conftime}
\end{equation}
since it absorbs all the information about the expansion of the universe (i.e., the scale parameter $a(\hat{t})$). In this case, the hierarchy of moment evolution equations \eqref{eq:Mmoment} becomes
\begin{equation}
\!\!\!\!\partial_\tau M_{n}(\tau \,)+M_{n}(\tau \,)=\frac{1}{n{+}1}%
\sum_{m=0}^{n}M_{m}(\tau \,)M_{n-m}(\tau \,).  \label{eq:KWmom}
\end{equation}
Interestingly enough, this equation  exactly coincides with the moment equation
originally derived by Bobylev~\cite{bobylev}, Krook, and Wu~\cite%
{KW-1,KW-2} for a non-relativistic, spatially homogeneous and isotropic,
non-expanding gas (see Eq.\ (35) in Ref.~\cite{KW-2}). The fact that the
non-equilibrium dynamics of both physical systems is governed by the same moment equations
is intriguing since the underlying symmetries of the two problems are quite
different. BKW's derivation is based on Galilean invariance while ours is
embedded into general relativity. One should note, however, that the
relation between the moments $M_{n}$ and the distribution function $f_k$
differs in the two cases: equations~\eqref{eq:rhomoments} and~%
\eqref{eq:defMmom} here are replaced in the non-relativistic case by
Eqs.~(18) and~(21) in Ref.~\cite{KW-2}.

Let us mention some important properties of the moment equations~\eqref{eq:KWmom}. First, the equation of motion~\eqref{eq:KWmom} implies that if $M_{n}(0)>0$ for all $n$ (which is true for any positive definite initial distribution function $f_k(0)$) then $M_{n}(\tau )$ will remain positive for $\tau \geq 0$. Equation~\eqref{eq:KWmom} shows that the $n^{\mathrm{th}}$ moment couples only to moments of the same or lower order. Therefore, for a given set of initial values for the moments $M_{n}(0)$ (or a given initial distribution function $f_k(0)$) we can express the solution $M_{n}(\tau )$ by a recursive procedure  in terms of the solutions $M_{m}(\tau )$ of lower order moments $m<n$. This can be seen explicitly by writing the general solution of Eq.~\eqref{eq:KWmom} formally as 
\begin{eqnarray}
&&M_{n}(\tau )=M_{n}(0)e^{-\omega _{n}\tau }\,  \label{eq:exactMnsol} \\
&&\quad +\,\frac{1}{n{+}1}\,\sum_{m=1}^{n-1}\,\int_{0}^{\tau }\,d\tau'
\,e^{\omega _{n}(\tau ^{\prime }{-}\tau )}\,M_{m}(\tau ^{\prime
})\,M_{n-m}(\tau ^{\prime })\,,  \notag
\end{eqnarray}
where 
\begin{equation}
\omega_{n}=1-\frac{2}{n+1}=\frac{n-1}{n+1}.  
\label{omegan}
\end{equation}
If rotational symmetry is broken, the evolution equation for $M_{n}$ includes additional couplings to moments of order $m>n$, rendering a recursive solution impossible~\cite{DeGroot:1980dk, Grad,Israel:1979wp, Denicol:2012cn, Muronga:2006zx}.

At sufficiently large times $\tau$ all moments $M_{n}(\tau )$ approach unity, independently of their initial condition. This can be seen explicitly by finding the fixed points of the set of equations~\eqref{eq:KWmom}, i.e., by studying the condition $\partial_{\tau }M_{n}(\tau )\bigl.\bigr|_{\tau =\tau _{\mathrm{max}}}=0,$ where $\tau_\mathrm{max}=\lim_{\hat{t}\rightarrow \infty }\tau \left( \hat{t}\right)$.
When imposing this condition on Eq.~\eqref{eq:KWmom} we obtain the following
recursion relation: 
\begin{equation}
M_{n}(\tau_\mathrm{max})=\frac{1}{n-1}\sum_{m=1}^{n-1}M_{m}(\tau _\mathrm{max})M_{n-m}(\tau
_\mathrm{max})\,.  \label{eq:fixpointMn}
\end{equation}%
This algebraic equation can be solved recursively as follows: the matching
conditions that define temperature and fugacity impose that $M_{0}(\tau
_\mathrm{max})=M_{1}(\tau _\mathrm{max})=1$. This gives immediately $M_{2}(\tau _\mathrm{max}){\,=\,}1$
for $n{\,=\,}2$. Induction shows that if $M_{m}(\tau _\mathrm{max}){\,=\,}1$ for $m<n$ then also $M_{n}(\tau _\mathrm{max}){\,=\,}1$ and, consequently, the stationary point of \eqref{eq:KWmom} is given uniquely by $M_{n}(\tau _\mathrm{max}){\,=\,}1$, for all values of $n$. In an FLRW universe, the Boltzmann equilibrium distribution is therefore the only fixed point of the Boltzmann equation. This analysis does not tell us whether or not the fixed point is an attractor; however, the validity of the H-theorem in FLRW spacetime \cite{Bernstein} necessarily guarantees that the equilibrium is stable. The numerical simulations reported below show that the equilibrium distribution is a stable (attractive) fixed point of Eq.~\eqref{eq:Boltzmanneq2}.

\subsection{Reconstructing the distribution function from Laguerre moments}
\label{subsec:lagmomenteqs} 

So far, the solution of the Boltzmann equation, $f_k(\tau)$, has been viewed as a function of the normalized energy moments $M_n(\tau)$. In practice it is, however, easier to reconstruct the distribution function from its moments if one uses a different set of moments, defined through a basis of orthogonal polynomials \cite{Grad}. In this paper we use the Laguerre basis \cite{Denicol:2012cn} (see Appendix \ref{app:Lague}) in which the distribution function can be written as
\begin{equation}
\label{eq:exactf}
  f_k(\tau) = f_k^{\mathrm{eq}}\,\sum_{n=0}^{\infty}\,c_{n}(\tau )\,\mathcal{L}_{n}^{(2)}
  \left( \frac{u{\,\cdot\,}k}{T(\tau )}\right) \,,
\end{equation}
where the Laguerre moments $c_n(\tau)$ are given by
\begin{eqnarray}
  &&\!\!\!\!\!\!c_{n}(\tau )=\frac{2}{(n{+}1)(n{+}2)}\,\frac{1}{\rho_0(\tau )}%
   \,\int_{k}\,(u\cdot k)\,\mathcal{L}_{n}^{(2)}\left( \frac{u{\,\cdot\,}k}{T(\tau)}\right) \,f_k\,\!  
\notag  \label{eq:Lagcoeff}\\
   &&\!\!\!\!\!\!=\sum_{r=0}^{n}\,(-1)^{r}\,\binom{n}{r}\,M_{r}(\tau )\,.
\end{eqnarray}
The second equality in this equation makes use of the closed form (\ref{eq:lagpoly}) of the Laguerre polynomials. For the Laguerre moments the particle number and energy conservation laws imply that 
\begin{equation}
c_{0}(\tau ){\,=\,}1,\quad c_{1}(\tau ){\,=\,}0\quad \mbox{for all }\tau
\label{c_cons}
\end{equation}%
(see Eq.~(\ref{M01})). The relation between $c_{n}$ and $M_{n}$ can be
inverted with the help of the binomial inverse transformation identity~\cite%
{riordan}: 
\begin{equation}
M_{n}(\tau )=\sum_{r=0}^{n}\,(-1)^{r}\,\binom{n}{r}\,c_{r}(\tau )\,.
\label{eq:mntocn}
\end{equation}%

In Appendix~\ref{app:nonlincn} we show that the Laguerre moments $c_{n}$
obey exactly the same hierarchy of coupled ordinary differential equations
as the normalized moments $M_{n}$: 
\begin{equation}
\partial_{\tau }c_{n}(\tau \,)+c_{n}(\tau \,)=\frac{1}{n{+}1}\sum_{m=0}^{n}c_{m}(%
\tau \,)c_{n-m}(\tau \,)\,.  \label{eq:cnmomeq}
\end{equation}
Only the initial conditions look different when expressed in terms of $M_n$ or $c_n$.

The structure of these equations has an interesting feature: since the right-hand side couples only to moments of lower order, one cannot generate low-order moments dynamically from higher-order ones. If initially all Laguerre moments up to order $n_\mathrm{min}$ vanish such that $c_{n_\mathrm{min}}$ is the lowest nonvanishing moment at $\tau=0$, it will remain the lowest nonvanishing moment at all times. This is useful when reconstructing the distribution function. Additionally, we note that the approach to thermal equilibrium $f_k \to f_k^{\mathrm{eq}} $ is characterized by $M_{n}(\tau ){\,\rightarrow \,}1$ for all $n$ and, consequently, $c_{n}(\tau ){\,\rightarrow \,}\delta _{n0}$. Also, using that $c_0(\tau)=1$ and $c_1(\tau)=0$ one finds that \eqref{eq:cnmomeq} can be rewritten as
\be
\partial_{\tau }c_{n}(\tau \,)+\omega_n c_{n}(\tau \,)=\frac{1}{n{+}1}\sum_{m=2}^{n-2}c_{m}(%
\tau \,)c_{n-m}(\tau \,)\,,  \label{eq:cnmomeqnew}
\ee
which will be useful in the next section when we discuss the linearized approximation for the collision kernel. 

Similar to the generic solution for the normalized energy moments $M_n$~\eqref{eq:exactMnsol}, for $n\geq 2$ Eq.~\eqref{eq:cnmomeqnew} admits a solution which for generic initial conditions reads as
\begin{eqnarray}
\label{eq:exactcnsolution} 
&&c_{n}(\tau )=c_{n}(0)e^{-\omega _{n}\tau }\,  
\\
&&\quad +\,\frac{1}{n{+}1}\,\sum_{m=2}^{n-2}\,\int_{0}^{\tau }\,d\tau'
     \,e^{\omega _{n}(\tau'{-}\tau)}\,c_{m}(\tau')\,c_{n-m}(\tau')\,,  
     \notag
\end{eqnarray}
The first term on the RHS corresponds to the linear contribution from the collision term ({\it cf.}  Eq.~\eqref{eq:lincnsol} in the following subsection) and decays exponentially with a rate $\omega_n$ that increases with $n$ according to Eq.~(\ref{omegan}). The non-linear second term describes the mode-by-mode coupling of $c_n$ with moments of lower order.  For small deviations from equilibrium ($c_n\ll1$ for all $n\ne0,1$) the linear, exponentially decaying terms dominate the dynamical evolution of the distribution function. For initially large deviations from equilibrium, however, no general statement  can be made as to which of the two terms (linear or nonlinear) controls the evolution at early times. Bounds on the nonlinear contribution to the generic solution of the Laguerre moments have been discussed for non relativistic systems~\cite{Cornille1980}. As we will see further below, at late times all $c_n$ eventually become small, and the remaining evolution is then controlled by the linear first term in Eq.~(\ref{eq:exactcnsolution}), i.e. if $\tau_l$ is large enough the moments $c_n$ relax for $\tau{\,>\,}\tau_l$ exponentially with rate $\omega_n$, $c_n(\tau{\,>\,}\tau_l)\approx c_n(\tau_l)\,e^{-\omega_n(\tau{-}\tau_l)}$.

In addition to the particle number and energy conservation laws $c_0(\tau)=1$ and $c_1(\tau)=0$, Eq.~\eqref{eq:exactcnsolution} yields the following exact solutions for the lowest order Laguerre moments (shown here up to $n\eq5$):
\begin{eqnarray}
\label{c2c3anal}
\!\!\!\!\!\!\!\!%
&&c_2(\tau) =  c_2(0)e^{-\omega_2 \tau},\qquad c_3(\tau) = c_3(0)e^{-\omega_3 \tau},
\nonumber \\ 
 \label{c4anal}
 &&c_4(\tau) =  c_4(0)e^{-\omega_4 \tau} + 3\, c_2^2(\tau) 
                         \left[ e^{-(\omega_4-2\omega_2) \tau}{-}1\right], 
 \\ \nonumber
&&c_5(\tau) = c_5(0)e^{-\omega_5 \tau} +  2\, c_2(\tau)c_3(\tau) 
\left[ e^{-(\omega_5-\omega_2-\omega_3)\tau}{-}1\right]\!.
\end{eqnarray}
One can see that mode-by-mode coupling among the Laguerre moments may start already at $n=4$, while for  $n<4$ the moments are either linear or completely determined by conservation laws. Another interesting feature of \eqref{eq:cnmomeq} is related to parity: if initially all the moments of $f_k$ with odd Laguerre polynomials vanish, $c_{2n+1}(0)\eq0$, the recursive nature of \eqref{eq:cnmomeq} implies that this remains true at all times: $c_{2n+1}(\tau)\eq0\ \forall\  t$. The same does not hold for initial conditions that have nonzero moments only with odd Laguerre polynomials. In this case, even Laguerre moments will in general be generated dynamically by mode-coupling between odd Laguerre moments, e.g. $c_6(\tau) = \frac{1}{2} c_3^2(\tau) \left(e^{2\tau/7}{-}1\right)$. This requires the full nonlinear collision term and hence does not happen when the latter is linearized as in the following subsection.

Finally, one sees from Eq.~\eqref{eq:exactf} that the distribution function at zero momentum $f(\tau,0)$ is finite at all times as long as the sum of the Laguerre moments remains finite. We will see later that at large times all Laguerre moments approach zero exponentially, rendering mode-coupling terms negligible for $\tau{\,\gg\,}1$. However, mode-coupling effects may be important if initial conditions are such that nonlinear terms, such as $c_2(\tau)^2 e^{-(\omega_4-2\omega_2) \tau}$ in \eqref{c4anal}, become of the same order as the linear contributions, in this case $\sim c_4(0)e^{-\omega_4 \tau}$. So while the moments decay exponentially at long times, their amplitudes in general still contain information about the nonlinear mode coupling at early times that cannot be obtained in linearized approaches such as the ones discussed in the next two subsections. 

\subsection{Moment evolution for a linearized Boltzmann collision term}
\label{subsec:linear} 

Systems not too far from local thermal equilibrium can be described
macroscopically using viscous hydrodynamics. To derive such hydrodynamic
equations from the underlying Boltzmann equation one expands the
distribution around the local equilibrium one, $f_k{\,=\,}{f_k^{\mathrm{eq}}}{+}%
\delta f_k$, and linearizes the Boltzmann equation in $\delta f_{k}$. When
representing the Boltzmann equation in terms of moments, this procedure
corresponds to a linearization of the moment equations around the
equilibrium values of the moments $M_{n}{\,=\,}1$ and $c_{n}{\,=\,}\delta
_{n0}$, respectively: 
\begin{eqnarray}
M_{n} &\approx &M_{n}^{\mathrm{lin}}=1+\delta M_{n}\,,   \label{clin}\\
c_{n} &\approx &c_{n}^{\mathrm{lin}}=\delta _{n0}+\delta c_{n}\,,
\end{eqnarray}%
with $\delta M_{0}{\,=\,}\delta M_{1}{\,=\,}\delta c_{0}{\,=\,}\delta c_{1}{%
\,=\,}0$ due to particle and energy conservation. The corresponding
linearized moment evolution equations, obtained from \eqref{eq:KWmom} and \eqref{eq:cnmomeqnew},
read 
\begin{eqnarray}
&&\partial_{\tau }\delta M_{n}(\tau )+\omega _{n}\delta M_{n}(\tau )=\frac{2}{n{+}1}%
\sum_{m=2}^{n-1}\delta M_{m}(\tau )\,,\qquad  \label{Mlineq} \\
&&\partial_{\tau }\delta c_{n}+\omega _{n}\delta c_{n}=0\,,
\label{clineq}
\end{eqnarray}
with $\omega _{n}$ given by Eq.~(\ref{omegan}). It is easy to check that these
linearized equations respect the relations \eqref{eq:Lagcoeff} and (\ref{eq:mntocn}), to linear order.

The general solution of Eq.~\eqref{Mlineq} is 
\begin{equation}
\begin{split}
\delta M_{n}& (\tau )=\delta M_{n}(0)e^{-\omega _{n}\tau }\, \\
& +\,\frac{2}{n{+}1}\,\sum_{m=2}^{n-1}\,\int_{0}^{\tau }\,d\tau ^{\prime}
e^{\omega _{n}(\tau ^{\prime }-\tau )}\,\delta M_{m}(\tau ^{\prime })\,,
\end{split}
\label{eq:sol-lineareq}
\end{equation}
while Eq.~(\ref{clineq}) is simply solved by 
\begin{equation}
\delta c_{n}(\tau )=e^{-\omega _{n}\tau }c_{n}(0).  \label{eq:lincnsol}
\end{equation}
These equations apply to moments with $n{\,\geq\,}2$. One sees that, in contrast to the linearized energy moments, the equations of motion for the linearized Laguerre moments decouple, i.e., the moments $\delta c_{n}$ are eigenfunctions of the linearized collision operator with eigenvalues (decay rates) $\omega _{n}$. The mode with the longest lifetime is the first non-hydrodynamic\footnote{%
    A non-hydrodynamic mode relaxes within a timescale that remains finite in homogeneous systems,
    in contrast to hydrodynamic modes such as sound waves.
    } 
mode, $n{\,=\,}2$, with $\tau _{2}{\,=\,}1/\omega _{2}{\,=\,}3$. As already noted, the decay rates increase with $n$, approaching unity for $n\rightarrow \infty $.

We can combine Eq.~\eqref{eq:lincnsol} with Eq.~\eqref{eq:mntocn} to obtain the following alternate solution of the linearized energy moments \eqref{eq:sol-lineareq}: 
\begin{equation}
\label{eq:sol-lineareq-2}
\delta M_{n}(\tau )=\sum_{r=2}^{n}(-1)^{r}\binom{n}{r}\,c_{r}(0)\,e^{-\omega
_{r}\tau }.  
\end{equation}
This form shows that, at asymptotically long times, the exponential decay of \emph{all} $M_{n}$ moments is controlled by $c_{n_\mathrm{min}}$, i.e. by the lowest initially non-vanishing Laguerre moment which has the smallest damping rate $\omega_{n_\mathrm{min}}$.

With the solution (\ref{eq:lincnsol}) of the linearized moment equations one finds the solution of the linearized Boltzmann equation\footnote{%
     In practice, in numerical calculations one truncates the infinite sums by defining a maximum 
     number of terms $n_{\mathrm{max}}$.
     } 
for the distribution function as follows (see Eq.\thinspace (\ref{eq:exactf})) :
\begin{equation}
\label{eq:exactlinear}
    f_k^{\mathrm{lin}}(\tau )=f_k^{\mathrm{eq}}
    \Bigl[1+\sum_{m=2}^{\infty}c_{m}(0)\,e^{-\omega _{m}\tau }\mathcal{L}_{m}^{(2)}\
           \Bigl(\frac{k}{T_{0}}\Bigr)\Bigr]\,
\end{equation}
%

\subsection{Moment evolution in the relaxation time approximation}
\label{subsub:rta} 

Due to its simplicity, one of the most widely employed models for the collision term is the Relaxation Time Approximation (RTA) \cite{BGK}. For our relativistic system it reads \cite{Anderson:1974} 
\begin{equation}
   \mathcal{C}[f]=-\frac{u\cdot k}{\tau _{\mathrm{rel}}(t)}\,\Bigl[f_k(t)-f_k^{\mathrm{eq}}\Bigr] \,,
\end{equation}
where $\tau _\mathrm{rel}$ is the scale at which the distribution function relaxes to its local equilibrium state. For the FLRW universe, the RTA Boltzmann equation is \citep{Bernstein, Cercignani} 
\begin{equation}
\label{eq:RTA-FLRW}
   \partial _{t}f_k(t)=-\frac{f_k(t)-f_k^{\mathrm{eq}}}{\tau _{\mathrm{rel}}(t)},  
\end{equation}
where according to Eq.~(\ref{eq:distreq-1}), $f_k^{\mathrm{eq}}$ is time independent. 

In general, the expression for the relaxation time $\tau _{\mathrm{rel}}$ varies according with the physical process one wants to investigate. For instance, the typical timescale for energy and momentum transport in the shear and bulk channels of relativistic fluids are in general different (this is the case in weak coupling QCD \cite{Arnold:2006fz}). A physical prescription must be given in order to meaningfully compare results compared within RTA and other evolution schemes. In this paper we choose to define $\tau _{\mathrm{rel}}$ in such a way that the shear viscosity to entropy density ratio of the gas, $\eta/s$, computed within RTA agrees with the result found using the full Boltzmann equation for massless particles with constant cross section \cite{Denicol:2012cn}. This condition fixes
\begin{equation}
\label{eq:taurFLRW}
   \tau _{\mathrm{rel}}(\hat{t})=\alpha\, a^{3}(\hat{t})\ell_0
\end{equation}  
with $\alpha = 1.58375$.\footnote{%
     The relaxation time was calculated from the Boltzmann equation for massless particles 
     interacting with constant isotropic cross section in Ref.~\cite{Denicol:2011fa, Denicol:2012cn}: 
     \begin{equation}
     \label{eq:tauest}
          \tau _{\mathrm{rel}}=\frac{5}{4}\frac{\eta }{\rho _{0}^{\mathrm{eq}}(t)\,T(t)}.  
     \end{equation}
     Here $\eta {\,=\,}1.267\,T/\sigma _{T}$ is the shear viscosity. Eq.~\eqref{eq:taurFLRW} is obtained 
     by using this value of $\eta$ together with $\rho _{0}^{\mathrm{eq}}(t)=n_{0}/a^{3}(t)$ in 
     (\ref{eq:tauest}).
     }
In this case, RTA Boltzmann equation becomes 
\begin{equation}
\label{eq:RTA-FLRW-2}
          \alpha \,\partial_{\tau}f_k(\tau )=f_k^{\mathrm{eq}}-f_{k}(\tau )\,,  
\end{equation}
which is easily solved analytically: 
\begin{equation}
\label{eq:RTAsol-1} 
          f_k^{\mathrm{RTA}}(\tau )=f_k^{\mathrm{eq}}+e^{-\tau/\alpha }\Bigl(f_k(0)-f_k^{\mathrm{eq}}\Bigr)\end{equation}
where $f_k(0)$ is the distribution function at $\tau {\,=\,}0$. Substituting this solution into the expression for the energy and Laguerre moments of the distribution function, we obtain the following analytic expressions for these quantities: 
\begin{eqnarray}
\label{eq:momentRTA}
c_{n}^{\mathrm{RTA}}(\tau )& =& c_{n}(0)e^{-\tau/\alpha }\,,  \label{eq:cnRTA} \\
M_{n}^{\mathrm{RTA}}(\tau )& =& 1+e^{-\tau/\alpha }\bigl(M_{n}(0)-1\bigr). \label{eq:normomRTA}
\end{eqnarray}
As before, we have $M_{0}^{\mathrm{RTA}}{\,=\,}{M}_{1}^{\mathrm{RTA}}{\,=\,}{%
c}_{0}^{\mathrm{RTA}}(0){\,=\,}1$ and $c_{1}^{\mathrm{RTA}}{\,=\,}0$ for all 
$\tau $, due to the conservation laws.

Whereas for the linearized collision term studied in the preceding subsection each moment $c_{n}$ relaxes with its own decay rate $\omega _{n}{\,=\,}(n{-}1)/(n{+}1)$, we see that the RTA collision term causes all of them to relax at the same decay rate $1/\alpha$. Since in the RTA the collision term is characterized by the single time scale $\tau _{\mathrm{rel}}$, this is perhaps not unexpected. On the other hand, linearizing the full Boltzmann collision term still leaves us with an infinite hierarchy of collision time scales, causing each of the Laguerre moments $\delta c_{n}$ \eqref{eq:lincnsol} to decay at its own rate $\omega_n$. One sees that by requiring $\eta/s$ computed in RTA to match the result from the full Boltzmann equation, the decay rate in RTA $1/\alpha \sim 0.631413$ comes to lie in between $\omega_4$ and $\omega_5$.   

We will show numerical comparisons between the solutions of the full nonlinear Boltzmann equation, its linearized form, and the RTA later in Sect.~\ref{sec:numres}. In the following section, however, we first use the methods developed in this section to rederive the exact analytical solution of the nonlinear Boltzmann equation presented in a previous publication~\cite{Bazow:2015dha}.

\section{An exact analytic solution of the Boltzmann equation with nonlinear collision kernel}
\label{sec:exact} 

\subsection{$\tau$-evolution of the distribution function and its moments}
\label{subsec:tau} 

One can see by inspection that Eq.~(\ref{eq:KWmom}) admits the following analytic solution of BKW type, valid for all $n\geq 0$ and $\tau\geq 0$ \cite{Bazow:2015dha}: 
\begin{eqnarray}
M_{n}(\tau ) &=&n\mathcal{K}^{n-1}(\tau )-(n{-}1)\mathcal{K}^{n}(\tau ),
\label{eq:kappasol} \\
\mathcal{K}(\tau ) &=&1\,-\,\frac{1}{4}\,e^{-\tau /6}\,.
\end{eqnarray}
Inserting the moments above into Eq.~\eqref{eq:Lagcoeff} gives the
corresponding analytic solution of Eq.~(\ref{eq:cnmomeq}) for the Laguerre
moments,\footnote{%
    This makes use of the combinatorial identity 
     \begin{equation}
      \sum_{r=0}^{n}\,(-1)^{r}\,\binom{n}{r}\,\bigl[rx^{r-1}-(r{-}1)x^{r}\bigr]\,=\,(1{-}n)\,(1{-}x)^{n}\,.
     \end{equation}%
} 
\begin{eqnarray}
c_{n}(\tau ) &=& (1{-}n)\bigl(1{-}\mathcal{K}(\tau )\bigr)^{n} = c_n(0)e^{-n\tau/6},
\label{eq:exactsolcn} 
\end{eqnarray}
where the initial values for the Laguerre moments are
\be
\label{CN}
c_{n}(0) =\frac{1-n}{4^{n}},
\ee
which can be obtained from the initial condition for the energy moments in \eqref{eq:kappasol}:
\be
\label{MN}
M_{n}(0) =\left( \frac{3}{4}\right) ^{n}\,\left( 1+\frac{n}{3}\right) \,.
\ee

In Ref.~\cite{Bazow:2015dha} we noted that the Fourier transform of the
distribution function can be expressed in terms of the normalized energy
moments and used this to construct the corresponding exact analytic solution
for the distribution function from $M_{n}$ as given in \eqref{eq:kappasol}.
This method is generalizable to any solution of the Boltzmann equation whose
Fourier transform exists. Here we rederive the same analytic solution for $%
f_k$ from the Laguerre moments, using the orthogonality and
completeness of the Laguerre polynomials. Inserting the analytic solution (%
\ref{eq:exactsolcn}) for the Laguerre moments into the decomposition (\ref%
{eq:exactf}) and using the relations \eqref{eq:lag-2} and \eqref{eq:lag-3}
listed in Appendix~\ref{app:Lague} we obtain %
\begin{eqnarray}
&&\!\!\!\!\!\!\!\!f_k(\tau )=f_k^{\mathrm{eq}}\,\sum_{n=0}^{\infty }\,(1{-}n)\left( 1{-}\mathcal{K}(\tau
)\right) ^{n}\mathcal{L}_{n}^{(2)}\left( k/T_{0}\right)
\label{eq:fullanboltzmann} \\
&=&\frac{\lambda \,e^{-k/(\mathcal{K}(\tau )T_{0})}}{\mathcal{K}^{4}(\tau )}%
\left[ 4\mathcal{K}(\tau )-3+\frac{k}{\mathcal{K}(\tau )T_{0}}\bigl(1{-}%
\mathcal{K}(\tau )\bigr)\right] .  \notag
\end{eqnarray}%
This agrees with Eq.~(22) in \cite{Bazow:2015dha}. This analytic solution of the nonlinear Boltzmann equation is obtained for the following far-from-equilibrium initial conditions for the distribution function (with energy and Laguerre moments given in (\ref{CN},\ref{MN})): 
\label{eq:ESini}
\begin{eqnarray}
   f_k(0) &=&\lambda \frac{256}{243}\,\left( \frac{k}{T_{0}}\right)
                    \exp \left( -\frac{4}{3}\frac{k}{T_{0}}\right) ,  \label{eq:ESICmn} 
\end{eqnarray}
%
%
\begin{figure*}[t]
\centering
\includegraphics[scale=0.21]{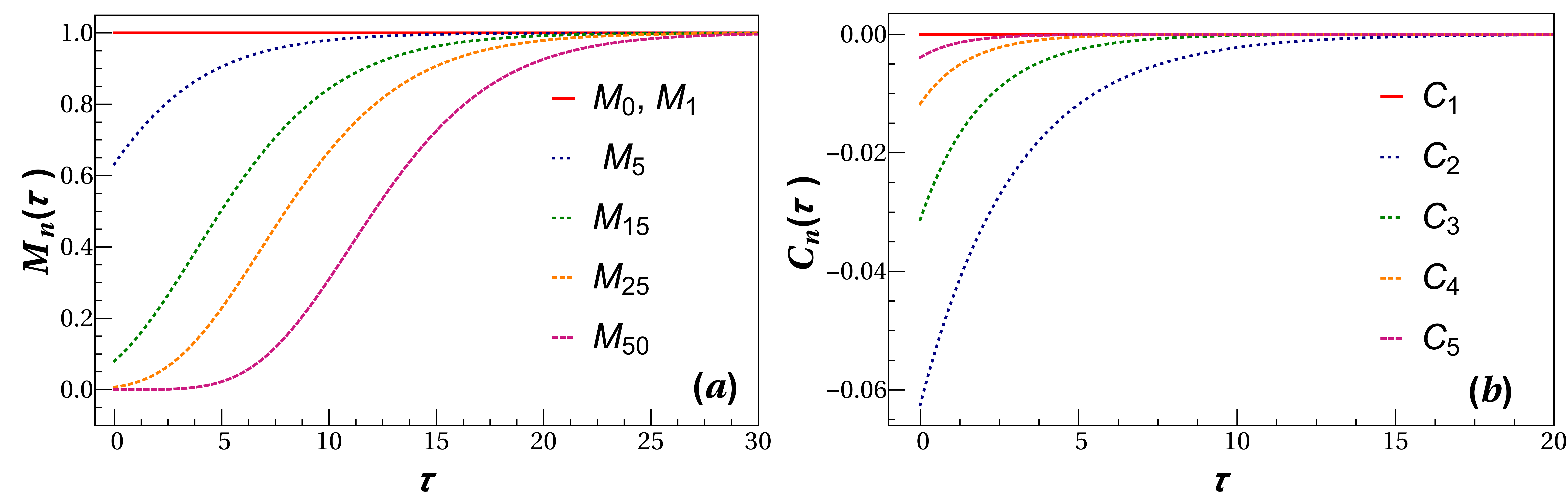}
\caption{{\protect\small (Color online) 
       Evolution of the normalized moments $M_{n}$~\eqref{eq:kappasol} (panel (a)) and the 
       Laguerre moments $c_{n}$ \eqref{eq:exactsolcn} (panel (b)) as a function of the dimensionless
       time variable $\tau$.}}
\label{F1}
\end{figure*}
%
The out-of-equilibrium initial condition in \eqref{eq:ESICmn} gives the opportunity to study how different approximations for the collision kernel affect the behavior of the Laguerre moments in an analytical manner. While all the evolution schemes correctly give exponentially decaying $c_n$'s, for the RTA solution in \eqref{eq:normomRTA} all the moments decay with the same scale $1/\alpha$ (given by the choice for the relaxation time). In the solution for the Laguerre moments obtained by linearizing the collision kernel in \eqref{eq:lincnsol} each moment decays exponentially at a distinct rate given by the eigenvalues of the collision operator, which should be in general a much better approximation to the multi timescale solution of the full nonlinear case in Eq.\ \eqref{eq:exactsolcn}. We note, however, that  for the exact solution studied in this section, the linearized collision kernel approach considerably underestimates the decay rates of the moments in which $n$ is large. This occurs because for the linearized moments \eqref{eq:lincnsol} one finds $\lim_{n\to \infty}c_n(\tau)/c_n(0)=e^{-\tau}$ while taking the same large $n$ limit in \eqref{eq:exactsolcn} gives $\lim_{n\to \infty}c_n(\tau)/c_n(0)=\lim_{n\to \infty}e^{-n\tau/6}\to 0$.\footnote{%
    Note that Eq.~(\ref{eq:exactsolcn}) implies that for this exact analytical solution all modes with
    $n{\,>\,}6$ decay faster than any of the eigenmodes $\omega_n$ of the linearized collision 
    operator, even at late times when all deviations from equilibrium $c_n$ ($n{\,\geq\,}2$) are small. 
    This exemplifies to the extreme the consequences of non-linear mode-coupling effects in the
    full Boltzmann collision operator. By linearizing the collision operator one loses essential information
    that is needed to describe correctly the dynamical evolution of the Laguerre moments for the exact 
    solution (\ref{eq:exactsolcn},\ref{CN}).
    }

%
\begin{figure}[b!]
\centering
\includegraphics[width=\linewidth]{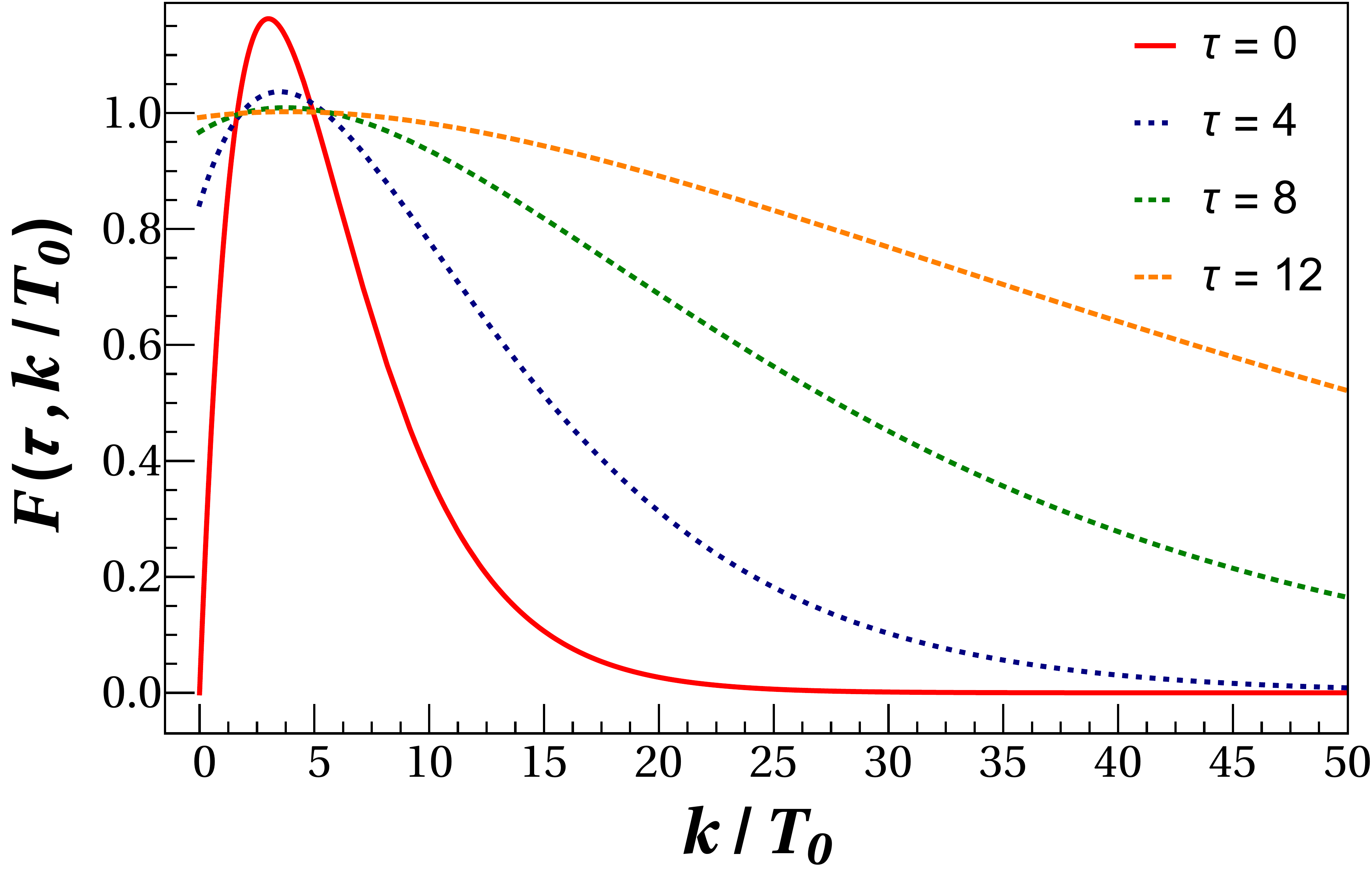}
\caption{(Color online) 
         Ratio between the out-of-equilibrium solution~\eqref{eq:fullanboltzmann} and the equilibrium 
         distribution as a function of $k/T_0$.
\label{F2}
}
\end{figure}
%

In Fig.~\ref{F1} we show the evolution of the the normalized moments $M_{n}$~\eqref{eq:kappasol} (left panel) and the Laguerre moments $c_{n}$~\eqref{eq:exactsolcn} (right panel) as a function of the dimensionless variable $\tau $. At $\tau =0$ the normalized moments $M_{n}$ decrease monotonically with increasing $n$. This means that the softest modes of the system are initially more strongly populated than the harder ones, albeit not thermally equilibrated (i.e. they are $<1$). The initial values for the Laguerre moments $c_{n}$ \eqref{eq:exactsolcn} are negative and increase (i.e. their magnitude decreases) with increasing $n$. Both $M_{n}$ and $c_{n} $ are seen to increase monotonically with time $\tau $ (i.e. the magnitudes of $c_{n}$ decrease monotonically), approaching their equilibrium values 1 and 0, respectively, at $\tau \rightarrow \infty $. 

This is more clearly seen in Fig.~\ref{F2} where we plot the ratio $F(\tau,k/T_{0})=f_k(\tau ) / f_k^{\mathrm{eq}}$ between the out-of-equilibrium solution~\eqref{eq:fullanboltzmann} and its equilibrium
value as a function of $k/T_{0}$ for different values of $\tau $. This ratio measures the deviation of the system from local equilibrium. At $\tau =0$ moderately soft modes with momenta $1.5\lesssim k/T_{0}\lesssim 5$ are overpopulated while the longest and shortest wavelength modes $k/T_{0}<1.5$ and $k/T_{0}>5$ are underpopulated. 

As time proceeds the distribution function approaches equilibrium: the
initial overpopulation at intermediate momenta quickly decreases, filling in
first the \textquotedblleft hole\textquotedblright\ at small momenta and
only later the strong initial depletion at large momenta. At $\tau {\,=\,}12$
the distribution function is seen to be essentially thermalized up to $k{%
\,\gtrsim \,}5T_{0}$, with a residual depletion of the high-momentum tail
that increases with $k$. 

Thermalization of the high-momentum modes appears
to require transporting energy from low to high momenta, similar to the
\textquotedblleft bottom-up\textquotedblright\ scenario in QCD \cite%
{Baier:2000sb} where interactions between the hard modes and the thermal
bath created by the soft modes allows the system to eventually reach global
thermal equilibrium asymptotically. The main difference between the QCD case
and the one at hand is that in the former the high-momentum modes are
initially over-occupied whereas here the initial conditions of the analytic
solution imply an initial under-population at high momenta. We will see in
Sec.~\ref{sec:numres} that the relatively slow thermalization of the
high-momentum part of the distribution function arises from
mode-by-mode-coupling effects characteristic of the non-linear Boltzmann
collision term with its broad spectrum of microscopic relaxation time
scales. This feature is not shared by the relaxation time approximation
where collisions are controlled by a single, common, relaxation
time.

\subsection{A finite $\tau$-horizon caused by cosmic expansion}

In the previous sections it was convenient to use the dimensionless variable 
$\tau$ (defined in Eq.~(\ref{eq:conftime})) as a time-like parameter in the
evolution equations for the moments $M_n$ and $c_n$. This was a key ingredient in demonstrating the relation between
our approach and the BKW solution. In this subsection we translate the
results obtained so far back into the original coordinate system, using the
time variable $t$.

The dynamics of the cosmic expansion is encoded in the scale factor $a(t)$
of the FLRW metric~\eqref{eq:FLRWds}. Its functional form is determined from
Friedmann's equation and depends on the equation of state~\cite%
{Weinberg-1,Weinberg-2,Bernstein}. For a conformal equation of state
consistent with our study of massless particles, the exact solution for the
scale factor $a(t)$ defined in Eq.~\eqref{eq:FLRWds} reads \cite%
{Weinberg-1,Weinberg-2}: 
\begin{equation}
a(\hat{t})=\sqrt{1+b_{r}\,\hat{t}}\;,\qquad b_{r}=\,2\,H_{0}\,\sqrt{\Omega _{r}},
\label{eq:scalarexp}
\end{equation}
where $H_{0}$ is the Hubble parameter evaluated at the initial time $\hat{t}_{0}=0$
and $\Omega _{r}$ is the dimensionless density parameter associated with
radiation. Equation~(\ref{eq:conftime}) then relates $\tau $ with $t$ as follows: 
\begin{equation}
\tau =\frac{2}{b_{r}}\,\left( 1-(b_{r}\,\hat{t}+1)^{-\frac{1}{2}}\right) \,.
\label{eq:tau2}
\end{equation}
This implies that the infinite $t$ interval $0\leq \hat{t}<\infty $ is mapped on a
finite $\tau $ interval $0\leq \tau \leq \tau_\mathrm{max}$ where 
\begin{equation}
\tau_\mathrm{max}=\lim_{\hat{t}\rightarrow \infty }\tau (\hat{t})=\frac{2}{%
b_{r}}.
\end{equation}

Consequently, the limit of perfect local thermalization of the distribution function at $\tau \rightarrow \infty $ is never reached: while (in contrast to the Gubser expansion studied in the context of relativistic heavy-ion collisions \cite{Denicol:2014tha, Denicol:2014xca, Gubser:2010ze}) a massless gas in equilibrium in FLRW spacetime would remain so despite the expansion of the universe, its cosmic expansion is not slow enough  to allow the system to ever reach complete local thermal equilibrium if it is initially out of equilibrium, $f_k(0) \neq f_k^{\mathrm{eq}}$. Instead, the system approaches a quasi-stationary off-equilibrium state characterized by the distribution 
\begin{eqnarray}
&&\lim_{\hat{t}\rightarrow \infty }f_k(\hat{t})=f_k(\tau _{\mathrm{%
max}})  
\label{eq:asympfun} \\
&&\quad =\frac{\lambda \,e^{-k/T_{\mathrm{lim}}}}{\mathcal{K}_{\mathrm{max}%
}^{4}}\left[ 4\,\mathcal{K}_{\mathrm{max}}-3+\frac{k}{T_{\mathrm{lim}}}\bigl(%
1{-}\mathcal{K}_{\mathrm{max}}\bigr)\right] .  \notag
\end{eqnarray}%
Here we defined the \textquotedblleft limiting
temperature\textquotedblright\ $T_{\mathrm{lim}}=\mathcal{K}_{\mathrm{max}%
}T_{0}$, with $\mathcal{K}_{\mathrm{max}}=1-\textstyle{\frac{1}{4}}\exp %
\left[ -\tau _{\mathrm{max}}/6\right]$.\footnote{%
    Recall that in terms of the physical momentum $k_\mathrm{phys}$ seen by a comoving observer 
    we have $k=a(t) k_\mathrm{phys}$ such that $k/T_\mathrm{lim}=k_\mathrm{phys}/T_\mathrm{lim}(t)$
    where $T_\mathrm{lim}(t)\equiv T_\mathrm{lim}/a(t)$ is the time-dependent (cosmologically redshifted)  
    ``limiting temperature'' seen by that observer. Since the temperature seen by the comoving observer 
    keeps redshifting we characterize the state (\ref{eq:asympfun}) as quasi-stationary.
    }
Figure~\ref{F2} shows that this spectrum is suppressed at high momenta relative to the asymptotic thermal distribution $\lambda \exp(-k/T_{0})$. The large-$k$ tail of the distribution is essentially exponential, with inverse slope parameter starting at $\textstyle{\frac{3}{4}}T_{0}$ at time $\hat{t}{\,=\,}0$ and increasing with time until it reaches $T_{\mathrm{lim}}=\mathcal{K}_{\mathrm{max}}T_{0}$ at time $\hat{t}\rightarrow \infty $. $\mathcal{K}_{\mathrm{max}}$ approaches 1 as the initial Hubble constant
(initial cosmic expansion rate) $H_{0}$ approaches zero.

\section{Entropy production by non-hydrodynamic modes}
\label{sec:entprod} 

The entropy density current $S^{\mu }$ defined in terms of the single particle distribution function in FLRW
\begin{equation}
\label{eq:entr-flow}
    S^{\mu }\equiv -\int_{k} k^{\mu }\,f_k\left( \ln f_k{-}1\right) 
\end{equation}
obeys Boltzmann's H-theorem \cite{Bernstein}, i.e. $d_\mu S^\mu \geq 0$, with the equality only being satisfied in equilibrium. Because of the symmetries of FLRW, one can write $S^\mu = s \,u^\mu$ with 
\be
   s=-\int_{k}(u{\,\cdot\,}k)\,f_k\,(\ln f_k{-}1)
\ee 
being the entropy density. Defining $\mathcal{S}=a^3 s$ and noticing that $a^6\,d_\mu S^\mu=\partial_\tau \mathcal{S}$, we find (using the decomposition \eqref{eq:exactf})
\bea
\partial_\tau \mathcal{S} &=& -\frac{n_0}{2}\sum_{n=2}^\infty h_n(\tau)\partial_\tau c_n(\tau) 
\label{eq:DA}
\eea
where 
\bea
h_n(\tau) = \int_0^\infty \!\!\!\! dx \,x^2 e^{-x} \mathcal{L}_n^{(2)}(x)
\ln\Bigl(1{+}\!\!\sum_{m=2}^\infty c_m(\tau)  \mathcal{L}_m^{(2)}(x) \Bigr).\ \
\label{eqan}
\eea

Equation (\ref{eq:DA}) shows that entropy production only ceases when the Laguerre moments become time independent, i.e., when equilibrium is reached. The lowest order moments $c_0$ and $c_1$, associated with hydrodynamic modes, do not participate in the entropy production which is entirely given by the non-hydrodynamic degrees of freedom $c_{n{\geq}2}$. In FLRW spacetime, local equilibrium is an attractor of the Boltzmann equation, i.e., a system initially prepared in local equilibrium will remain in local equilibrium (in spite of the cosmological expansion) while an initially non-equilibrated system will evolve towards local equilibrium, producing entropy along the way. What is different from other situations is that the evolution of the macroscopic hydrodynamic observables such as the energy and particle number densities follows the laws of ideal fluid dynamics even if the system is out of equilibrium. This happens because, in the present situation which has an exceptional degree of symmetry, the non-hydrodynamic modes completely decouple from the energy momentum tensor, thereby preserving its ideal fluid form. Similar systems were studied before in \cite{Noronha:2015jia,Hatta:2015kia}.

In dissipative fluid dynamics entropy production is expressed in terms of the non-equilibrium corrections to the energy-momentum tensor and particle 4-current. For example, in the widely used Israel-Stewart formulation of dissipative fluid dynamics \cite{Israel:1979wp}, entropy production is expressed in terms of the shear stress tensor $\pi^{\mu \nu }$ as $d_\mu S^{\mu }=\pi ^{\mu \nu }\pi _{\mu \nu }/(2\eta T) $, where $\eta $ is the shear viscosity. The shear stress tensor reflects the excitation of non-hydrodynamic modes of the Boltzmann equation \cite{Denicol:2011fa,Denicol:2012cn}, and such a formulation is expected to work if the system is sufficiently close to thermodynamic equilibrium. However, the entropy production derived in \eqref{eq:DA} can obviously never be expressed in a hydrodynamic form even if the system is close to equilibrium. Therefore, the type of system discussed here gives an example in which the symmetries of the system always forbid the description of dissipative processes (such as entropy production) in terms of the laws of fluid dynamics.

Based on the discussion above, one can use \eqref{eq:cnmomeqnew} to find another expression for the entropy production
\be
\partial_\tau \mathcal{S} = \frac{n_0}{2}\biggl[\sum_{n=2}^\infty \omega_n\,h_n c_n - 
\sum_{n=2}^\infty  \frac{h_n}{n{+}1} \sum_{m=2}^{n-2} c_{n-m}c_m\biggr].
\label{eq:DAnew}
\ee
This expression shows that, at late times when the Laguerre moments $c_n$ are small, entropy production in the full nonlinear case (in which the moments follow \eqref{eq:cnmomeqnew}) should be very well approximated by the corresponding expression computed in the linearized Boltzmann collision approximation. In fact, one can expand the logarithm in \eqref{eqan} to linear order to find
\be
h_n(\tau) = (n{+}1)(n{+}2)\, c_n(\tau) +\mathcal{O}(c_n^2)
\label{approximatehn}
\ee
and, thus,
\be
    \partial_{\tau}\mathcal{S} = \frac{n_0}{2}\sum_{n=2}^\infty (n{+}1)(n{+}2)\,\omega_n c_n^2(\tau) 
     + \mathcal{O}(c_n^3).
\label{newentropyproduction}
\ee
In this limit the fact that entropy increases with time is manifest, and each moment is seen to contribute to entropy production an amount proportional to its decay rate. If the higher order corrections in Eq.~\eqref{newentropyproduction} are small (as they are for the initial conditions studied in this work), one expects that the linearized Boltzmann collision approximation should give an accurate description of the entropy produced in the full nonlinear problem. This is confirmed in the numerical studies performed in Section \ref{subsec:entr-prod-comp}. Also, using \eqref{approximatehn} one finds that 
\be
\mathcal{S}(\tau) = \mathcal{S}_{\mathrm{eq}} - \frac{n_0}{4}\sum_{n=2}^\infty (n{+}1)(n{+}2)\,c_n^2(\tau) + \ldots
\ee
where $\mathcal{S}_{\mathrm{eq}}$ is the corresponding equilibrium expression. This shows explicitly that the maximum entropy value is achieved in equilibrium.

Equation~(\ref{eq:DAnew}) expresses the entropy production in terms of the time evolution of the Laguerre moments of the distribution function. In Sec.~\ref{sec:FLRWmethod} we studied their evolution for the full nonlinear Boltzmann collision terms as well as for a linearized version and for the relaxation time approximated collision term. In the next section we will show numerical results for these different types of microscopic evolution. Following the production of entropy in each of these three cases will yield valuable insights into the dynamics that underlies the thermalization processes in the Boltzmann equation.

\section{Numerical results}
\label{sec:numres} 

%
\begin{figure*}[t]
\centering
\includegraphics[width=0.8\textwidth]{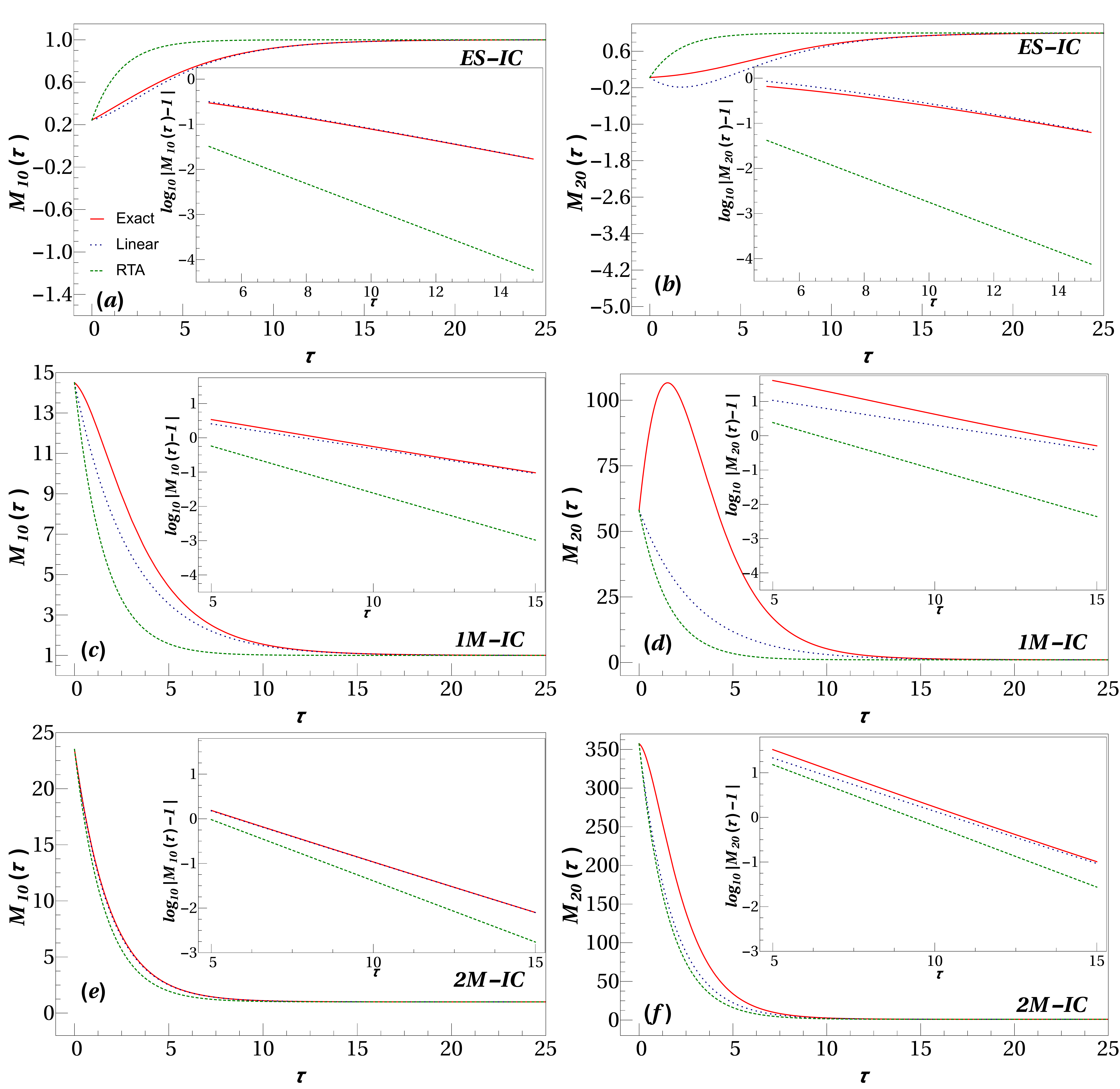}
\caption{(Color online) 
        Evolution of the moments $M_{10}$ (left column) and $M_{20}$ (right column) as a 
        function of the dimensionless time $\protect\tau$ according to the nonlinear evolution 
        equation \eqref{eq:KWmom}, its linearized version \eqref{Mlineq}, and within the RTA 
        \eqref{eq:normomRTA} for the ES-IC~\eqref{eq:ESini2} (panels (a)-(b)), 
        1M-IC~\eqref{eq:1Mini} (panels (c)-(d)), and 2M-IC \eqref{eq:2Mini} (panels (e)-(f)). 
        See text for further details.
\label{F3}
}
\end{figure*}
%

In this section we compare the solutions to the full nonlinear Boltzmann equation \eqref{eq:exactf}, its linearized version~\eqref{eq:exactlinear} and the RTA~\eqref{eq:RTA-FLRW}. For simplicity we assume that the fugacity is $\lambda=1$. We consider the following initial conditions of the distribution function $f_k(0)$, the Laguerre moments $c_n(0)$, and the normalized moments $M_n(0)$:
\begin{itemize}
\item The exact solution initial condition (ES-IC) already given in Eqs.~(\ref{eq:exactsolcn})-(\ref{MN}): 
\begin{subequations}
\label{eq:ESini2}
\begin{eqnarray}
f_k(0)&=&\lambda\frac{256}{243}\,\left(\frac{k}{T_0}\right)\, e^{-\frac{4}{3}\frac{k}{T_0}}\,, \\
c_n(0)&=&\frac{1-n}{4^n}\,, \\
M_n(0)&=&\left(\frac{3}{4}\right)^{n}\,\left(1+\frac{n}{3}\right)\,.
\end{eqnarray}
\end{subequations}

\item The one mode initial condition (1M-IC): 
\begin{subequations}
\label{eq:1Mini}
\begin{eqnarray}  
f_k(0)&=&\lambda\,e^{-\frac{k}{T_0}}\left[1+\frac{3}{10}\mathcal{L
}_2^{(2)}\left(\frac{k}{T_0}\right)\right]\,, \\
c_n(0)&=&\delta_{n0}+\frac{3}{10}\,\delta_{n2}\, \\
M_n(0)&=& 1+\,\frac{3}{10}\,\binom{n}{2}.
\end{eqnarray}
\end{subequations}

\item The two-mode initial condition (2M-IC): 
\begin{subequations}
\label{eq:2Mini}
\begin{eqnarray}  
f_k(0)&=&\lambda\,e^{-\frac{k}{T_0}}\left[1-\frac{1}{10}\,%
\mathcal{L}_3^{(2)}\left(\frac{k}{T_0}\right)\,\right. \\
&&\qquad\qquad\ \left.+\, \frac{1}{20}\,\mathcal{L}_4^{(2)}\left(\frac{k}{T_0%
}\right)\right]\,, \\
c_n(0)&=&\delta_{n0}-\frac{1}{10}\,\delta_{n3}+\frac{1}{20}\,\delta_{n4}\,,
\\
M_n(0)&=& 1+\frac{1}{10}\,\binom{n}{3}\,+\,\frac{1}{20}\,\binom{n}{4}.
\end{eqnarray}
\end{subequations}
\end{itemize}

All of these initial conditions satisfy the requirement $M_n(0)\geq 0$ for all $n$, which ensures positivity of the distribution function $f_k$. Notice also that for 1M-IC~\eqref{eq:1Mini} and 2M-IC~\eqref{eq:2Mini} the normalized moments $M_n(0)$ diverge when $n \to \infty$, which should be contrasted with the ES-IC case\eqref{eq:ESICmn} where $M_n(0)$ vanishes in this limit.  

In this section we will compare for each of these initial conditions the
evolution of the moments of the distribution function, the phase-space
evolution of the distribution function reconstructed from the Laguerre
moments, and the amount of entropy produced in the evolution, for the evolution schemes defined by the full nonlinear Boltzmann collision kernel, its linearized version, and also the
relaxation time approximation.

\subsection{Evolution of the moments}
\label{subsec:comp-func} 

\begin{figure*}[tbp]
\begin{centering}
\includegraphics[width=0.9\textwidth]{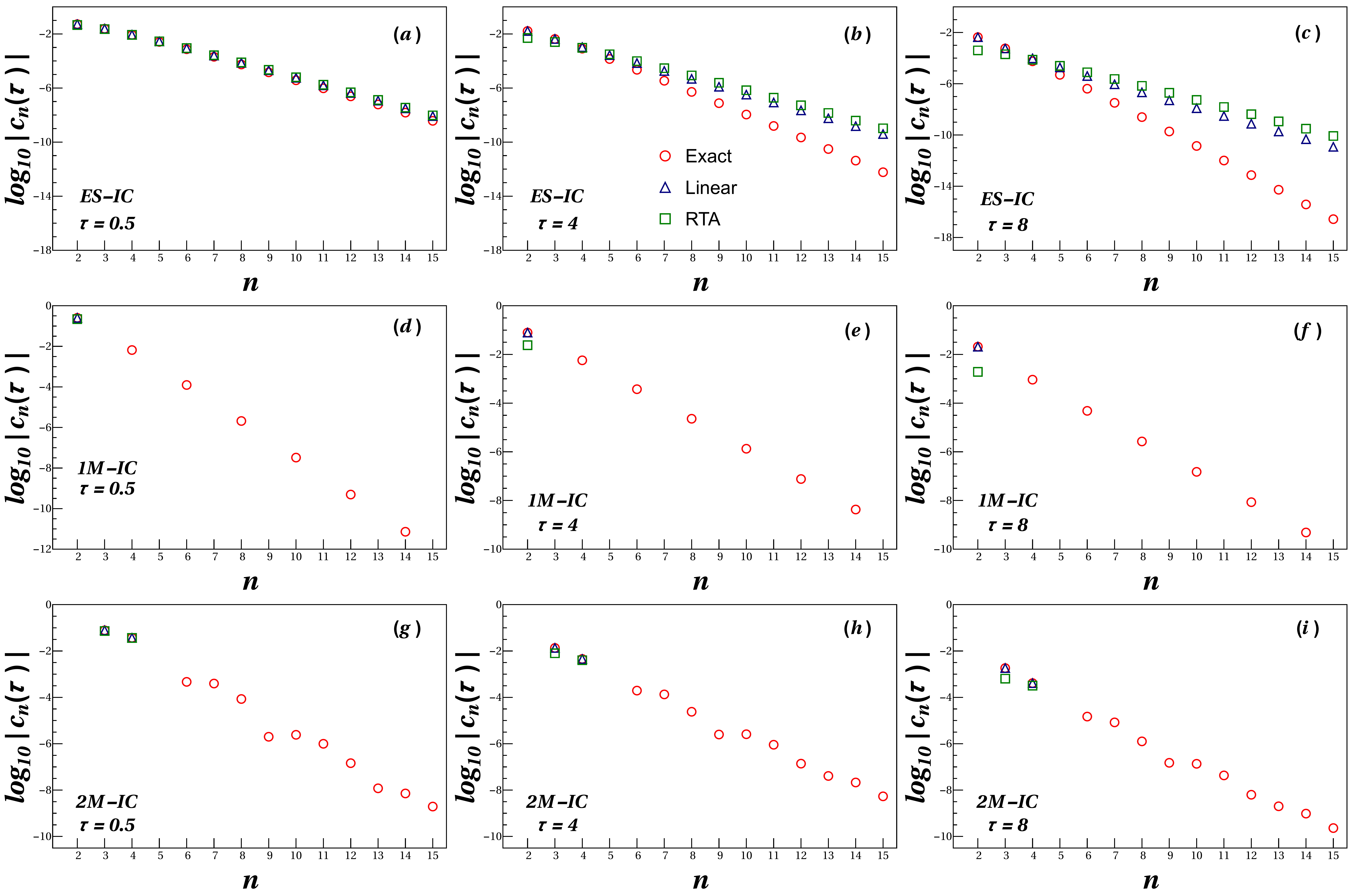}
\end{centering}
\caption{(Color online) 
        Evolution of the Laguerre moments as a function of $n$ according to the nonlinear Boltzmann 
        equation (red circle), linear approximation (blue triangle) and RTA (green square) for fixed 
        values of $\tau=\{0.5,4,8\}$ (left, middle and right column respectively). For the initial conditions 
        of the distribution function we use the ES-IC~\eqref{eq:ESini2} (panels (a,b,c)), 
        1M-IC~\eqref{eq:1Mini} (panels (d,e,f)), and 2M-IC~\eqref{eq:2Mini} (panels (g,h,i)).
\label{F4}
}
\end{figure*}

The infinite set of differential equations~\eqref{eq:KWmom} for the
normalized moments $M_{n}$ is truncated at a finite $n_{\mathrm{max}}$ and
then solved numerically. The evolution of the linearized moments $%
M_{n}^{\mathrm{lin}}=1+\delta M_{n}$ is obtained by solving the differential
equations \eqref{Mlineq} for $\delta M_{n}$, with initial conditions fixed
by $\delta M_{n}(0)=M_{n}(0)-1$. Within the RTA the evolution
of the moments $M_{n}^{\mathrm{RTA}}$ is determined by Eq.~%
\eqref{eq:normomRTA}. 

Figure~\ref{F3} shows the numerical solutions for the moments $M_{10}$ (left column) and $M_{20}$ (right column) as functions of the dimensionless variable $\tau$ for the three initial conditions listed above. We observe that the difference between the values of the moments $M_{n}$, $M_{n}^{\mathrm{lin}}$, and $M_{n}^{\mathrm{RTA}}$ gets larger as one increases the order $n$ of the moment. Since higher-order moments are more strongly weighted at higher momenta, these differences indicate that the RTA and the linear approximation of the Boltzmann collision term provide descriptions of the microscopic dynamics that degrade at short distance scales.

At large times, all $M_{n}$ moments relax exponentially to their equilibrium value of 1. In the insets in Fig.~\ref{F3} we plot the difference between $M_{n}(\tau )$ and their asymptotic value on a logarithmic scale, in order to visualize the rate of approach to equilibrium of each moment. We see that in
RTA the moments $M_{n}^{\mathrm{RTA}}$ relax much faster to their asymptotic
value than for both the full and linearized Boltzmann collision terms. In RTA all modes relax exponentially at the same rate $\omega =1/\alpha$ (defined by our choice for the relaxation time) while for both the
full and linearized Boltzmann collision term the energy moments $M_{n}$ mix
Laguerre moments of different orders that decay with different rates $\omega
_{n}<1$. At large times $\tau $, their decay is dominated by the moment with
the smallest decay rate, namely the first non-vanishing non-hydrodynamic
mode $n_{\mathrm{min}}$. For ES-IC and 1M-IC, the lowest non-vanishing
non-hydrodynamic mode is $c_{2}$, and correspondingly for both the full and
linearized Boltzmann collision terms the $M_{n}$ modes decay asymptotically
with $\omega _{2}{\,=\,}1/3$. For 2M-IC the lowest non-vanishing
non-hydrodynamic mode is $c_{3}$, and correspondingly for both the full and
linearized Boltzmann collision terms the $M_{n}$ modes decay asymptotically
somewhat faster, with $\omega _{3}{\,=\,}1/2$. We also note that at large
times the deviations of the moments from their asymptotic values become
small, and the time evolutions of the linearized and full moments converge.

At early times the faster relaxation of the $M_n^\mathrm{RTA}$ moments to their equilibrium values compared to their relaxation for the full and linearized collision term is most evident.
However, Fig.~\ref{F3} also shows that for some of the initial conditions
the early-time evolution of the $M_n$ moments exhibits dramatic differences
even between the full and linearized collision terms. These differences
arise from mode coupling effects which are generically large as long as the
moments $M_n$ deviate strongly from their equilibrium values.

\begin{figure*}[t!]
\begin{centering}
\includegraphics[width=\textwidth]{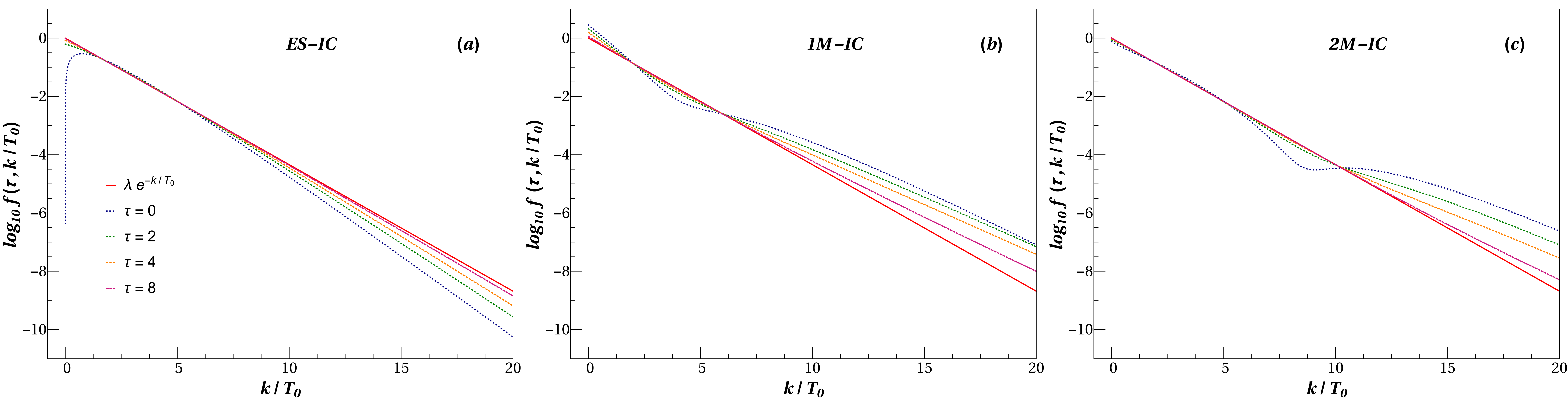}
\end{centering}
\caption{(Color online)
     Snapshots of the full nonlinear distribution function as a function of $k/T_0$ at different values 
     of $\tau=\{0,2,4,8\}$ (with fugacity $\lambda=1$). For the initial conditions of the distribution 
     function we use the ES-IC~\eqref{eq:ESini2} (panel a), 1M-IC~\eqref{eq:1Mini} (panel b), and
     2M-IC~\eqref{eq:2Mini} (panel c).
     \label{F5}
}
\end{figure*}

We point out that for the initial condition ES-IC the moment $M_{20}^\mathrm{lin}$ becomes negative at early times, specifically in the interval $0\lesssim \tau\lesssim 7$ (see panel (b) in Fig.~\ref{F3}).\footnote{%
     We have checked numerically that for $20{\,\leq\,}n{\,\leq\,}100$ all the
     moments $M_n^\mathrm{lin}$ turn negative somewhere in the interval $\tau\in (0,7)$.
     } 
From their definition it is clear that this cannot happen for a distribution function that is positive definite. This dynamical behavior resulting from the linearization of the moments around their thermal equilibrium values is thus unphysical. We will see later that, for the initial conditions of the exact solution discussed in Sec.~\ref{sec:exact}, this unphysical dynamics causes the distribution function to turn negative at large momenta as time proceeds, somewhat reminiscent of a similar phenomenon observed for the exact solution of the RTA Boltzmann equation in a system undergoing Gubser expansion \cite{Denicol:2014tha,Heinz:2015cda}.

In Fig.~\ref{F4} we present the logarithm of the magnitude of the Laguerre moments $|c_n|$ as a function of $n$ for a set of fixed $\tau$ values, $\tau=\{0.5,4,8\}$ for the left, middle and right column, respectively, and for the initial conditions mentioned above. As shown in Eqs.\,\eqref{c2c3anal}, the solutions for the moments with $n\leq 3$ are the same in the nonlinear case as in the linearized Boltzmann approximation, and this is observed in Fig.~\ref{F4}. For the ES-IC at early times there is basically no distinction among the different evolution schemes (since the initial $c_n$ are already all nonzero while nonlinear mode-coupling effects have not yet had a chance to manifest themselves). As time evolves the $c_n$'s with large $n$ quickly distinguish nonlinear evolution (red circles) from linear evolution schemes (denoted by the blue triangles and green squares);  however, only at late times can clearly distinguish (especially at large $n$) between the results from the RTA and the linearized Boltzmann approach. 

For the 1M-IC only $c_0$ and $c_2$ are initially nonzero (parity even), and one can see that $c_{2n+1}(\tau)=0$ for all $\tau$, as explained in Sec.~\ref{subsec:lagmomenteqs}. Also, this case clearly shows the effect of mode-by-mode coupling responsible for exciting for $\tau>0$ modes with $n>2$ even though they were initially zero,. This should be contrasted with the linear evolution schemes that give $c_{n>2}(\tau)=0$
for all $\tau$. The difference in the decay rate for $c_2$ between the RTA and the nonlinear (and linearized) case is evident in panel f of Fig.\ \ref{F4}. 

Since $c_2(0)=0$ for 2M-IC, Eq.~(\ref{c4anal}) implies that $c_4(\tau)$ obtained in the nonlinear evolution is identical to the result computed within the linearized Boltzmann approximation, i.e. $c_4(\tau) = c_4(0)e^{-\omega_4 \tau}$. This can be seen in Fig.~\ref{F4}g,h,i. Once again mode-by-mode coupling in the full nonlinear evolution is responsible for exciting for 2M-IC moments with $n>4$ which become nonzero already after a short time $\tau =0.5$. The higher moments can only be excited by nonlinear coupling with lower modes which is an effect not included in either the RTA or the linear Boltzmann approximation. 

The Laguerre moments $\{ c_n\}$ contain all the information about the solutions of the Boltzmann equation.
However, the way $\{ c_n\}$ encodes this information is not at all trivial. For instance, in the deep infrared $f(\tau,k{\,\to\,}0) = f_k^\mathrm{eq} \sum_{n=0}^\infty c_n(\tau)$. Also, the complicated time evolution of the analytical solution in Eq.\ \eqref{eq:fullanboltzmann} is translated into a simple exponential decay of the Laguerre modes described by \eqref{eq:exactsolcn}.   

\subsection{Evolution of the distribution function}
\label{subsec:linvsnonlin} 
\vspace*{-2mm}

\begin{figure*}[t!]
\begin{centering}
\includegraphics[width=\textwidth]{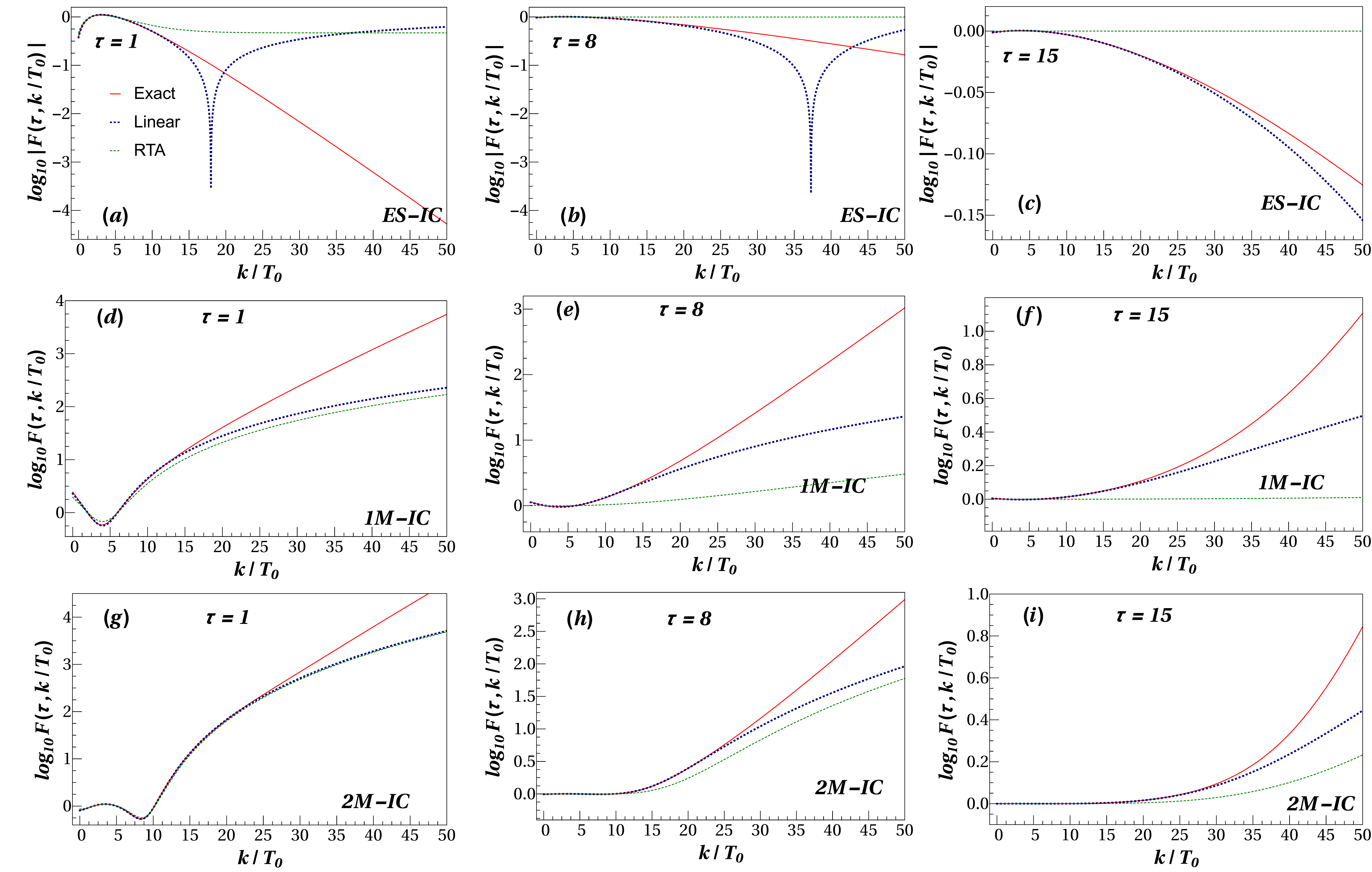}
\end{centering}
\caption{(Color online) 
      Snapshots of the ratio $F(\tau,k/T_0){\,\equiv\,}f_k(\tau)/f_k^\mathrm{eq}(\tau)$ as a 
      function of $k/T_0$ for $\protect\tau=\{1,8,15\}$ (left, middle and right column) according 
      to the nonlinear Boltzmann equation (red line), linear approximation (blue dashed line) 
      and RTA (green dotten line). For the initial conditions of the distribution function we use 
      the ES-IC~\eqref{eq:ESini2} (panels (a,b,c)), 1M-IC~\eqref{eq:1Mini} (panels (d,e,f)), 
      and 2M-IC~\eqref{eq:2Mini} (panels (g,h,i)). 
\label{F6}
}
\end{figure*}

The recursive structure of the evolution equations \eqref{eq:cnmomeq}, \eqref{eq:lincnsol} and \eqref{eq:normomRTA} for the Laguerre moments of the distribution function makes it easy to systematically improve the description of the distribution function by increasing the truncation order $n_\mathrm{max}$ (the total number of moments) until convergence is achieved.\footnote{%
      For the initial conditions studied in this work we were able to ensure convergence of the series 
      \eqref{eq:exactf} at all times with a small number $n_\mathrm{max}$ of associated Laguerre 
      polynomials. However, for other (still well-behaved) initial conditions (e.g. a Gaussian bump 
      added to a thermal distribution) the polynomials $\mathcal{L}^{(2)}_n$ are not well-adapted 
      to describe the high-momentum tail of the distribution function, and we found it necessary to 
      include a very large number $n_\mathrm{max}$ of these polynomials to ensure convergence 
      of the series for $f_k(\tau)$.
      }

\vspace*{-2mm}
\subsubsection{Evolution of non-thermal energy tails}
\label{subsub:form} 
\vspace*{-2mm}

In Sec.~\ref{sec:exact} we saw for the exact analytical solution \eqref{eq:fullanboltzmann} of the full Boltzmann equation that, while it  approaches equilibrium at large $\tau$, hard momenta are being occupied very slowly and large deviations from equilibrium persist for high values of $k/T_0$ at very large $\tau$. Here we study numerically how $f_k(\tau)$ evolves towards equilibrium with the full nonlinear Boltzmann collision term for the two other initial conditions listed at the beginning of this section and compare it with the evolution of the ES-IC initial condition for which we have an exact analytic result.

Figure~\ref{F5} shows that the other initial conditions correspond to initial distribution functions which deviate from equilibrium even more strongly than the one corresponding to the exact solution, albeit in different momentum regions. In these initial conditions hard modes are separated from soft modes by a ``kink" (located, e.g., near $k/T_0=9$ for 2M-IC) that is more distinct in the 2M-IC case than in the 1M-IC (where it also occurs at a lower value of $k/T_0$). Taking all three panels of the figure together one observes that the low-momentum region $k/T_{0}{\,\lesssim \,}5$ relaxes to equilibrium very quickly, reducing deviations from equilibrium occupancy to below 20\% already at $\tau \sim 2$ while at $k/T_{0}{\,>\,}20$ deviations from equilibrium by up to a factor 5 persist up to $\tau \sim 10$. The Boltzmann collision terms thus thermalizes the system differentially: first the system reaches approximate equilibrium at thermal length scales whereas thermalization at sub-thermal length scales takes much longer.

\vspace*{-2mm}
\subsubsection{Comparing the evolution of the distribution function for
different approximations of the collision kernel}
\label{subsub:comp} 
\vspace*{-2mm}

\begin{figure*}[t!]
\begin{centering}
\includegraphics[width=0.8\textwidth]{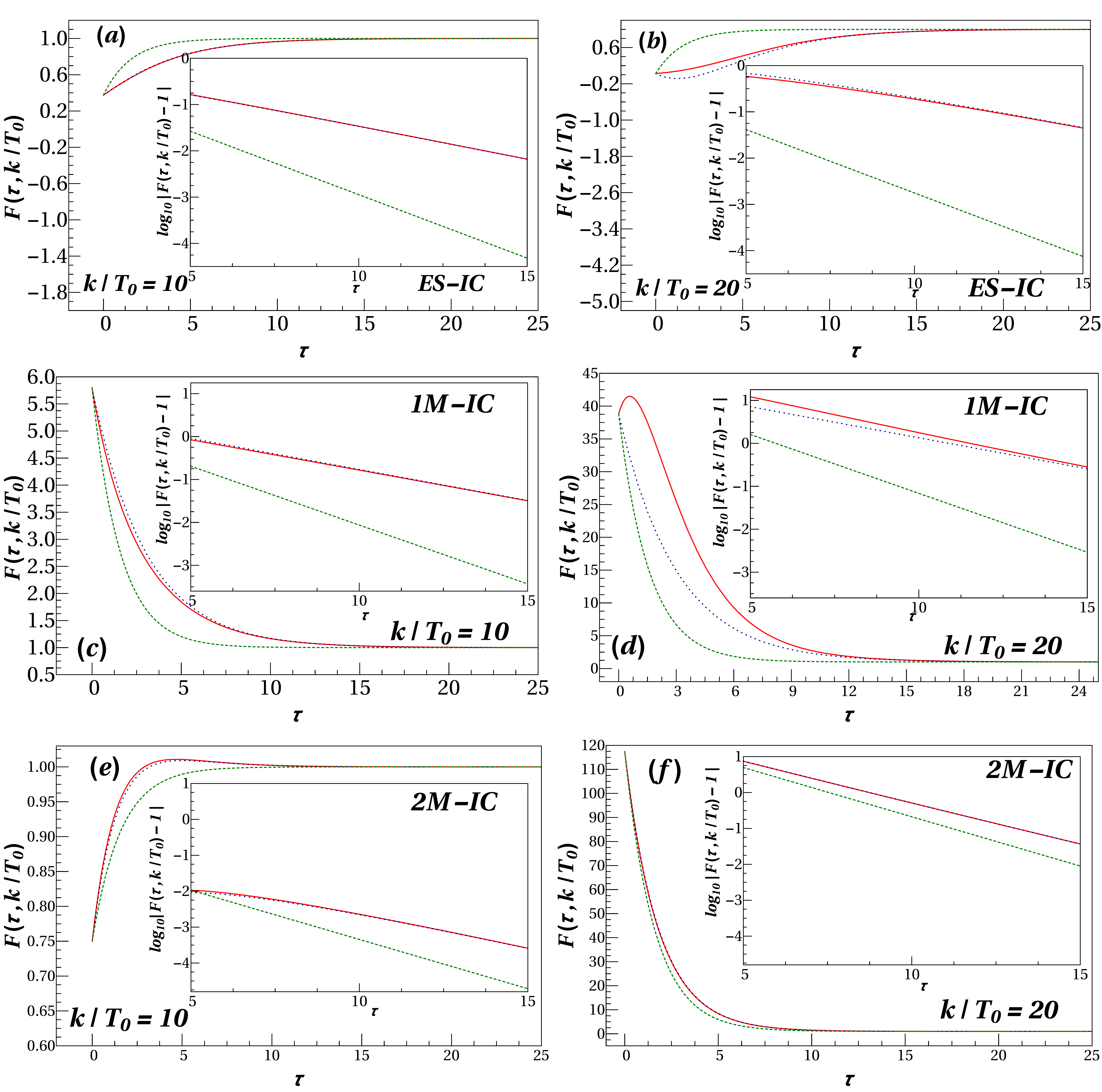}
\end{centering}
\caption{(Color online) 
(Color online)  Evolution of the ratio $F(\tau,k/T_0){\,\equiv\,}f_k(\tau)/f_k^\mathrm{eq}(\tau)$ as a
function of $\tau$, for fixed values of momentum $k/T_0=10$ (left column) and $k/T_0=20$ (right column), for the full nonlinear (red
line), linear (dotted blue line) and RTA collision term (green dotten line). For the initial conditions we use the ES-IC~\eqref{eq:ESini2}
(panels (a,b)), 1M-IC~\eqref{eq:1Mini} (panels (c,d)), 2M-IC~\eqref{eq:2Mini} (panels (e,f)).}
\label{F7}
\end{figure*}

In Figs.~\ref{F6} and~\ref{F7} we compare numerical results for the phase-space evolution of the distribution function for the full nonlinear solution to the Boltzmann equation~\eqref{eq:exactf}, its linear approximation~\eqref{eq:exactlinear}, and the RTA~\eqref{eq:RTAsol-1} for all three sets of initial conditions. In Fig.~\ref{F6} we plot the logarithm of the ratio $F(\tau,k/T_0){\,\equiv\,}f_k(\tau)/f_k^\mathrm{eq}(\tau)$ of the non-equilibrium distribution function to its equilibrium value as a function of $k/T_0$, for a set of fixed $\tau$ values, $\tau=\{1,8,15\}$ (left, middle, and right column). In the first row, we plot the magnitude $|F|$ of this ratio because, for the linearized collision term, the distribution function evolves to negative values at large momenta as time proceeds. This behavior was already anticipated in Sec.~\ref{subsec:comp-func} where we saw that some of the energy moments $M_n$ became unphysically negative when evolved with the linearized evolution equations. Fig.~\ref{F6}a, b, c shows that the pathological region of negative distribution 
%
\begin{figure*}[t!]
\begin{centering}
\includegraphics[width=\textwidth]{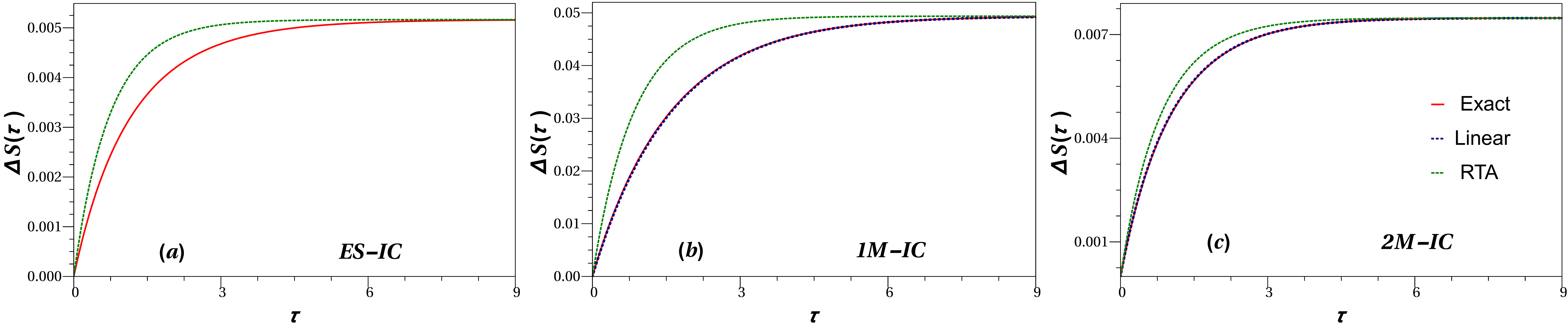}
\end{centering}
\caption{(Color online) 
        Time evolution of the produced entropy as a fraction of its initial value, 
        $\Delta\mathcal{S}(\tau)$ as defined in Eq.~(\ref{eq:totalentr}), for initial conditions 
        (a) ES-IC \eqref{eq:ESini2}, (b) 1M-IC~\eqref{eq:1Mini}, and (c) 2M-IC~\eqref{eq:2Mini}.
\label{F8}
}
\end{figure*}
%
functions appears to move to larger momenta as time proceeds. This is consistent with the observation that at large $\tau$ the deviations from equilibrium get smaller and the linear approximation to the full Boltzmann collision term (which does not cause the distribution function to become negative) thus can be expected to work better. We note once again that with our choice for the relaxation time the RTA evolved distribution function reaches equilibrium much more quickly than both the nonlinear case and the linearized Boltzmann collision approximation; the slowest approach to equilibrium is observed when the system is evolved with the full nonlinear collision term. The three lower rows of panels further show that the momentum range in which the dynamically evolved distribution function closely approaches equilibrium grows wider, extending to larger momenta as time proceeds.

Figure~\ref{F7} shows the time evolution of the same ratio $F$ plotted in \ref{F6} at two different momenta ($k/T_0=10$ and 20, respectively). Similar to what we saw for the evolution of
its energy moments, one observes a rapid approach to thermal equilibrium in the
RTA evolution, compared to the much slower thermalization found using the nonlinear 
collision term. At the lower of the two selected $k/T_0$ values, differences
between the time evolution for the full and the linearized Boltzmann collision
term are hardly noticeable. For the larger $k/T_0{\,=\,}20$, the early-time
evolution differs significantly between the full nonlinear and the linearized
collision terms for ES-IC and 1M-IC; in particular, for the exact analytic
solution the linearized time evolution leads to unphysical negative values of
the distribution function at early times.

At late times, the difference between the value at a given momentum of the evolving non-equilibrium distribution function and its thermal limit decreases exponentially. The rate of approach to equilibrium is $\omega\eq1/\alpha$ in RTA, as is expected because all its Laguerre moments decay exponentially with this rate. For the full Boltzmann collision term, the thermalization rate converges at late times to $\omega _{2}\eq1/3$ for ES-IC and 1M-IC and to $\omega _{3}\eq1/2$ for 2M-IC, i.e. at large times thermalization is controlled by the lowest (and slowest) non-vanishing non-hydrodynamic moment (which is $n\eq2$ for ES-IC and 1M-IC and $n\eq3$ for 2M-IC). This asymptotic late-time behavior is universal in the sense that it applies at all momenta.

\subsection{Entropy production}
\label{subsec:entr-prod-comp} 

We quantify the total entropy produced during the thermalization process by the fractional increase
\begin{equation}  \label{eq:totalentr}
\begin{split}
\Delta\mathcal{S}(\tau)&=\frac{\mathcal{S}(\tau)-\mathcal{S}(0)}{\mathcal{S}%
(0)}\,.
\end{split}%
\end{equation}
The time evolution of $\Delta\mathcal{S}$ is studied in Fig.~\ref{F8} for the three initial conditions for the distribution function listed at the beginning of this section.\footnote{%
       We do not show the entropy production for the linearized evolution of the initial conditions 
       ES-IC, Fig~\eqref{F8}a, since this leads to negative distribution functions in part of momentum 
       space for which the entropy integral is not defined.
       }
All cases have the same initial energy and particle density which evolve according to ideal fluid dynamics to the same final equilibrium state at $\tau\to\infty$. What is different in each case is the initial entropy of the system. The different initial conditions correspond to non-equilibrium configurations and, thus, their initial entropy is lower than the equilibrium value. Since equilibrium is a global attractor of the dynamics, the relative difference
\be
    \Delta \mathcal{S}_{\mathrm{eq}}\equiv 
    \frac{\mathcal{S}_{\mathrm{eq}}-\mathcal{S}(0)}{\mathcal{S}_{\mathrm{eq}}}
\ee
gives the amount of entropy produced over all time for each initial condition. We find $\Delta \mathcal{S}_{\mathrm{eq}}=0.51\%$ for ES-IC, $\Delta \mathcal{S}_{\mathrm{eq}}=4.7\%$ for 1M-IC, and $\Delta \mathcal{S}_{\mathrm{eq}}=0.74\%$ for 2M-IC. Thus, we see that 1M-IC is the initial condition that is the farthest from equilibrium and, consequently, produces the largest amount of entropy during the evolution.   

As should be expected from the thermalization studies of the distribution function and its moments in the preceding subsections, the initial rate of entropy production and the approach of the total entropy towards its final equilibrated value is fastest in the relaxation time approximation. When the kinetic evolution is controlled by the full or linearized Boltzmann collision term, the rate of entropy production slows down to an asymptotic exponential approach at the rate $\omega _{n_{\mathrm{min}}}{\,=\,(}n_{\mathrm{min}}{-}1)/(n_{\mathrm{min}}{+}1)$, where $n_{\mathrm{min}}$ is the order of the lowest initially non-zero Laguerre moment of $f_k$. In Fig.~\ref{F7} one can clearly distinguish between the different rates towards thermal equilibrium between panel (b) where initially the lowest nonzero non-hydrodynamic moment is $c_{2}$ which relaxes to equilibrium with the rate $\omega _{2}{\,=\,}1/3$, and panels (c,d) where initially the lowest nonzero non-hydrodynamic moment is $c_{3}$, which in turn relaxes to equilibrium at a faster rate $\omega _{3}{\,=\,}1/2$.

As expected from the discussion in Section \ref{sec:entprod}, there are no noteworthy differences in the entropy production rate for the full and the linearized Boltzmann collision term. High momentum particles are too rare to significantly contribute to the total entropy of the system, which means that long before the high-momentum tails of the distribution function become thermal the overall entropy production has already essentially ceased. In other words, the total entropy is dominated by particles with thermal momenta, and entropy production essentially stops when those thermal particles have reached an equilibrium state.

\vspace*{-2mm}
\section{Conclusions}
\label{sec:concl} 
\vspace*{-2mm}

In this work we solved the full nonlinear Boltzmann
equation for an expanding massless gas with constant cross section in FLRW
spacetime. The problem of solving the nonlinear Boltzmann equation is mapped onto solving
recursively a set of coupled ordinary differential equations of moments of
the distribution function. The precision of the solution can be improved
systematically to any desired value by increasing the number of moments
(which results in better resolution of the high-momentum tail of the
distribution function). The same method can be applied to the Boltzmann
equation with a linearized collision term or using the relaxation time
approximation (RTA), which allowed us to investigate the importance for the
thermalization process of non-linear mode-by-mode couplings inherent in the
Boltzmann collision term.

The cosmological expansion in FLRW spacetime was found to be slow enough to allow the distribution function to move towards local equilibrium for any initial condition. Local equilibrium is reached in the asymptotic limit $\tau\to\infty$ when expressed in the dimensionless time variable $\tau$ defined in Eq.~(\ref{eq:conftime}). This asymptotic limit can, however, only be reached for FLRW universes with infinitesimally small initial values of the Hubble constant. For finite initial expansion rates, the limit $t\to\infty$ in physical time is reached after a finite interval in $\tau$, which leaves the distribution function in a non-equilibrium final state that becomes approximately stationary at late physical times.

Our work exhibited a characteristic difference between the rates at which the system approaches thermal equilibrium in RTA and for the full or linearized Boltzmann collision term. For both the full and linearized collision terms, the asymptotic thermalization rate for the distribution function is $\omega_{n_\mathrm{min}}\eq\frac{n_\mathrm{min}{-}1}{n_\mathrm{min}{+}1}$, which is the damping rate of the \emph{slowest} initially occupied non-hydrodynamic eigenmode $n_\mathrm{min}$ of the Boltzmann equation. In RTA, on the other hand, if the relaxation time is calculated with standard methods using the same (constant) cross section as in the Boltzmann collision term, the distribution function approaches equilibrium at the larger rate $\omega\eq1/\alpha$ which falls between $\omega_4$ and $\omega_5$.

The approach to equilibrium is fastest for typical thermal momenta whereas the high-momentum tail of the distribution function takes much longer to thermalize. As time proceeds the window in which the distribution is well approximated by the asymptotic equilibrium distribution widens towards larger momenta. The late thermalization of the high energy tails is caused by non-linear mode-by-mode couplings that couple higher moments to lower ones and transport energy from low to high momenta. It is consistent with the simple intuitive picture that high-momentum particles require multiple collisions to thermalize whereas soft particles thermalize already after a few collisions \cite{KW-2}.

Although the dynamics generated by the full collision term
exhibits non-linear mode-coupling effects, we found only very small
differences in the evolution towards equilibrium between the full and
linearized Boltzmann collision terms as long as we restricted our attention
to the dominant thermal momentum region; significant differences between the
linear and nonlinear thermalization dynamics were, however, observed at
large momenta or short length scales. Since high-momentum particles
contribute very little to the total entropy of the system, the rate of
entropy production during the thermalization process was found to be almost
indistinguishable between the nonlinear and linearized dynamics. In RTA,
however, entropy was was found to be produced at much higher rate, leading
to faster thermalization.

An interesting aspect of this model is that it combines ideal fluid
dynamical evolution with dissipation and entropy production. This means that
the rate of entropy production cannot be expressed in the standard way
through dissipative flows (such as the shear stress tensor), which vanish in
our model exactly by symmetry. Dissipative effects, while definitely
present, do not manifest themselves hydrodynamically, i.e., they do not
affect the (relatively slow) evolution of the hydrodynamic modes whose
dynamics is controlled by the conservation laws. For a given initial
particle and energy density, the amount of entropy produced depends
exclusively on the how far the initial phase-space distribution is away from
thermal equilibrium; all initial configurations with the same particle and
energy density eventually evolve to the same equilibrium state at $\tau\to\infty$.

The dramatically different thermalization time scales for the RTA and full
nonlinear Boltzmann collision terms raise the question whether one could
not simply bring the Boltzmann equation in RTA in congruence with the full
nonlinear Boltzmann equation by appropriate ``renormalization'' of the
relaxation time $\tau_\mathrm{rel}$ used in the RTA. However, this does not
work: as our analysis shows, for the full Boltzmann collision term the
relaxation time towards thermal equilibrium is not universal, but depends on
which of the non-hydrodynamic modes are initially occupied. Thermalization
happens asymptotically at the rate $\omega_{n_\mathrm{min}}$ where $n_%
\mathrm{min}$ is the order of the slowest initially non-zero
non-hydrodynamic mode. Using in RTA a relaxation time that depends on the
initial condition for the distribution function (i.e. on which
non-hydrodynamic Laguerre moments are initially non-zero) does not make
sense.

It will be interesting to try to extend the techniques developed in this
work to physically interesting anisotropically expanding systems. Finding an
exact solution of the full nonlinear Boltzmann equation for
(0+1)-dimensional Bjorken~\cite{Bjorken:1982qr} and/or the (1+1)dimensional
Gubser \cite{Gubser:2010ze} flows in Minkowski space would be of particular
practical and conceptual interest for relativistic heavy-ion physics, and in
the cosmological context one would like to be able to solve the Boltzmann
equation in anisotropic spacetimes such as the Bianchi universes \cite{Krasinski}. We leave these issues for future studies.

\section*{Acknowledgements}

We gratefully acknowledge inspiring discussions with S.~Schlichting, S.~K\"onig, Y.~Mehtar-Tani, S.~Ozonder and A.~Dumitru. MM thanks C.~Plumberg for pointing out Ref.~\cite{riordan} on combinatorial identities. JN thanks Conselho Nacional de Desenvolvimento Cient\'{\i}fico e Tecnol\'{o}gico (CNPq) and Funda\c{c}\~{a}o de Amparo \`{a} Pesquisa do Estado de S\~{a}o Paulo (FAPESP) for financial support. UH and MM express their gratitude to the Institute for Nuclear Theory, where part of this work was realized, for
its hospitality. GSD was supported by DOE Contract No. \rm{DE-SC0012704}. DB, UH and MM are supported by the U.S. Department of Energy, Office of Science, Office for Nuclear Physics under Award DE-SC0004286. UH, MM, and JN acknowledge support through a bilateral travel grant from FAPESP and The Ohio State University.\\

\appendix
\section{Moments of the collision kernel}

\label{app:mom-collker} 

In this Appendix we describe the procedure to perform the integrals~%
\eqref{eq:coll-mom} for the case of a constant cross section. We start
by calculating the term $\mathcal{C}_{loss}^{(n)}$~\eqref{eq:loss1}. We recall
from Sec.~\ref{sec:FLRWspace} that we formulate all momentum integrals in
terms of the \emph{covariant} components of the four-momenta. For massless particles in FLRW spacetime we have $s=(k{+}%
k^{\prime })\cdot(k{+}k^{\prime })=2\,\left(1{-}\cos\theta\right)\,k\,k^{%
\prime 2}$ and $u \cdot k =k^0=k/a(t)$.

The term $\mathcal{C}_{loss}^{(n)}$~\eqref{eq:loss1} is calculated as
follows 
\begin{eqnarray}  
\label{eq:loss-s1}
\mathcal{C}_\mathrm{loss}^{(n)}&=&\frac{(2\pi)^5}{2}\,\sigma_T\,\int_{kk^{%
\prime }pp^{\prime }}\!\!\!\! s\,(u{\cdot}k)^n\,  \notag \\
&&\times\sqrt{-g} \,\delta^{4}(k{+}k^{\prime }{-}p{-}p^{\prime })\,f_k f_{k^{\prime }}\,,  \notag \\
&=&\frac{\sigma_T}{a^{n+2}}\,\int_{kk^{\prime }} k^{n+1}\,k^{\prime
}\,\left(1{-}\cos\theta\right)\,f_k f_{k^{\prime }} \\
&=&\sigma_T\,\left[\frac{1}{2\pi^2}\frac{1}{a^{n+3}}\int_0^\infty
dk\,k^{n+2}\,f_k \right]  \notag \\
&&\times\,\left[\frac{1}{4\pi^2}\frac{1}{a^3}\int_0^\infty dk^{\prime
}\,k^{\prime 2}\, f_{k^{\prime }} \int_{-1}^1d(\cos\theta) (1{-}%
\cos\theta)\right]  \notag \\
&=&\sigma_T\,\rho_n \,\rho_0\,.  \notag
\end{eqnarray}
In the second line we used the identity 
\begin{equation}
\int_{pp^{\prime }}\sqrt{-g}\,\delta^{4}(k{+}k^{\prime }{-}p{-}p^{\prime
})=\frac{1}{(2\pi)^5}\,,
\end{equation}
and in the last equality we recalled the definition \eqref{eq:rhomoments} of
the energy moments. This completes the computational details of the loss
term $\mathcal{C}_\mathrm{loss}^{(n)}$~\eqref{eq:loss1}.

The calculation of $\mathcal{C}_{gain}^{(n)}$~\eqref{eq:gain2} is harder.
Let us start by rewriting Eq.~\eqref{eq:gain2} as follows 
\begin{equation}  \label{eq:gain-s1}
\mathcal{C}_\mathrm{gain}^{(n)}=\frac{(2\pi)^5}{2}\,\sigma_T\,\int_{kk^{%
\prime }}\,s\,f_k f_{k^{\prime }}\,\mathcal{P}_n\,,
\end{equation}
where we define the scalar quantity 
\begin{equation}
\mathcal{P}_n = \int_{p\,p^{\prime }}\,(u{\cdot}p)^{n} \,\sqrt{-g}\,
\delta^{4}(k{+}k^{\prime }{-}p{-}p^{\prime })\,.  \label{eq:Pn}
\end{equation}
We calculate $\mathcal{P}_n$ in the center of mass frame where $\mathbf{k}{+}\mathbf{k^{\prime }}{\,=\,}0$, where $\mathbf{k}=(k_1,k_2,k_3)$ is constructed from the covariant spatial components of the 4-vector $k^\mu$. (The same applies to all 3-vectors below, see discussion in Sec.~\ref{sec:FLRWspace}.) In this frame the total energy of the system is $\sqrt{s}=k^0+k^{\prime 0}$. In this reference frame the fluid velocity is not static, i.e. $u^\mu=(u^0,u^i)$ has nonzero spatial components. $\mathcal{P}_n$ is then calculated as follows: 
\begin{widetext}
\be
\label{eq:Pn2}
\begin{split}
    \mathcal{P}_n &=\frac{1}{(2\pi)^6}\,\int\frac{d^3p}{\sqrt{-g}\,p^0}\,(u{\cdot}p)^{n}
    \int\frac{d^3p'}{\sqrt{-g}\,p'^{0}}\, \delta\left(\sqrt{s} - \frac{1}{a(t)}\left(p + p'\right)\right)\,
    \delta^{3}(\mathbf{p}+\mathbf{p'}) \,,\\
&= \frac{1}{2\,(2\pi)^5}\,\frac{1}{a^{n+1}(t)}\,\int_0^\infty dp \,p^{n}\delta\left(\sqrt{s}-2\frac{p}{a(t)}\right)\,\int_0^\pi d\theta \sin\theta\, \,\left( u_0- \frac{|\mathbf{u}|}{a(t)}  \cos \theta    \right)^{n} \,,\\
&=\frac{1}{2^{n+1}\,(2\pi)^5}\,\frac{a(t)}{(n+1)\,\sqrt{s}\,|\mathbf{u}|}\,
\left[\left(u_0\sqrt{s}+\frac{|\mathbf{u}|\sqrt{s}}{a(t)}\right)^{n+1} -\left(u_0\sqrt{s}-\frac{|\mathbf{u}|\sqrt{s}}{a(t)}\right)^{n+1}\right]\,.
\end{split}
\ee
The last expression can be written covariantly by introducing the total 4-momentum of the system $P^\mu = k^\mu + k'^\mu$ such that
\be
 u_0\sqrt{s} = u{\cdot}P \,,\qquad
 P{\cdot}P\equiv P^2 = s \,,\qquad
 \sqrt{s}|\mathbf{u}| = \sqrt{s u_0^2 -s} = \sqrt{(u{\cdot}P)^2 - P^2}\,. 
\ee
Thus the covariant version of Eq.~\eqref{eq:Pn2} is 
\be
\label{eq:Pn3}
\mathcal{P}_n = \frac{1}{2^{n+1}\,(2\pi)^5}\,\frac{a(t)}{(n+1)\,\sqrt{(u{\cdot}P)^2{-}P^2}}
\left[\left(u{\cdot}P+ \frac{\sqrt{(u{\cdot}P)^2 - P^2}}{a(t)}\right)^{n+1}
-\left(u{\cdot}P-\frac{\sqrt{(u{\cdot}P)^2 - P^2}}{a(t)}\right)^{n+1}\right]\,,
\ee
Next we use the identity
\be
\frac{(x{+}y)^{n+1}-(x{-}y)^{n+1}}{y} = 2 \sum_{\substack{r=1\\r \mathrm{odd}}}^{n+1} \,\bin 
\,x^{n+1-r}\, y^{r-1}
\ee
to write Eq.~\eqref{eq:Pn3} as 
\be
\label{eq:Pn4}
\mathcal{P}_n = \frac{1}{(n+1)\,(2\pi)^5\,2^n}\,
\sum_{\substack{r=1\\r \in\, odd}}^{n+1} \,\bin\,(u{\cdot}P)^{n+1-r}\,a(t)^{1-r}
\left[(u{\cdot}P)^2 - P^2\right]^{(r{-}1)/2}\,.
\ee
In the fluid rest frame one has $u{\cdot}P=P^0=(k+k')/a(t)$ and $(u{\cdot}P)^2-P^2={\bf P}\cdot{\bf P}=({\bf k}+{\bf k'})^2$. Thus, the scalar $\mathcal{P}_n$ finally reads
\be
\label{eq:finalpn}
\begin{split}
\mathcal{P}_n  = \frac{1}{(n+1)\,(2\pi)^5\,2^n}\frac{1}{a^n(t)}\sum_{\substack{r=1\\r \in\, odd}}^{n+1}\bin\,(k+k')^{n+1-r}  \left| \mathbf{k}+\mathbf{k}' \right|^{r-1}\,.
\end{split}
\ee
Substituting the last expression back into Eq.~\eqref{eq:gain-s1} we obtain
\be
\label{eq:gain-s2}
\begin{split}
\mathcal{C}_\mathrm{gain}^{(n)}&=\frac{\st}{(2\pi)^6}\,\frac{1}{a^{6+n}(t)}
\int\,d^3 k\,d^3 k'\,\left(1-\cos\theta\right)\,f_kf_{k'} 
\frac{1}{(n+1)\,2^{n+1}}\,\sum_{\substack{r=1\\r \in\, odd}}^{n+1} \,\bin\,(k+k')^{n+1-r}|\mathbf{k}+\mathbf{k'}|^{r-1} \,,\\
&=\frac{\st}{(2\pi)^6}\,\frac{1}{a^{6+n}(t)}\,\int\,d^3 k\,d^3 k'\,\left(1-\cos\theta\right)\,f_kf_{k'} \\ 
&\times  \frac{1}{(n+1)\,2^{n+1}}\,\sum_{r=0}^{n+1} \,\bin\,(k+k')^{n+1-r}(k^2+k'^2+2k\,k'\,\cos\theta)^{(r-1)/2}\,\left[1+(-1)^{r+1}\right]\,.
\end{split}
\ee
To perform the integration over the angular variable $\theta$ we need the following integral:
\be
\begin{split}
\int_{-1}^1 dx\, (1-x)\, \left(a\,+\,b\, x\right)^{\frac{r-1}{2}} =\,4\frac{(a+b)^{(r+3)/2}-(a-b)^{(r+1)/2} \left(a+b (2+r)\right)}{b^2\,(r+1)\,(r+3)}\,,
\end{split}
\ee
valid as long as $a\geq b$ and $b\geq 0$. Eq.~\eqref{eq:gain-s2} then reads
\be
\label{eq:gain-s3}
\begin{split}
\mathcal{C}_\mathrm{gain}^{(n)}=& \frac{\st}{a^{6+n}(t)}\,\frac{1}{2^{n+2}}\,
\,\sum_{r=0}^{n+1} \,\binom {n}{r} \frac{\left[1+(-1)^{r+1}\right]}{(r+1)(r+3)}\int_0^\infty \frac{dk}{2\pi^2} \int_0^\infty \frac{dk'}{2\pi^2}\,f_kf_{k'} \\ 
&\times  \,\left\{ (k+k')^{n+4}-|k-k'|^{r+1}(k+k')^{n+1-r}\left[k^2+k'^2+2k\,k' (2+r)\right]
\right\}\,,\\
= & \frac{2\st}{a^{6+n}(t)}\,\frac{1}{(n+1)(n+3)(n+4)}\,\int_0^\infty \frac{dk}{2\pi^2}\int_0^\infty \frac{dk'}{2\pi^2}f_{k} f_{k'}  \\ 
&\times \left\{ (k+k')^{n+4}-k^{n+4} -k^{n+3}k' (n+4)-k'^{n+4} -k'^{\,n+3}k(n+4) \right\} 
\end{split}
\ee
where we used the following identities:
\begin{subequations}
\begin{align}
&\frac{1}{2^{n+2}}\sum_{r=0}^{n+1}\,\binom {n}{r} \frac{\left[1+(-1)^{r+1}\right]}{(r+1)(r+3)} =\frac{2^{n+3}-n-5}{2^{n+2}(n+1)(n+3)(n+4)}\,,\\
&\frac{1}{2^{n+2}}\sum_{r=0}^{n+1} \,\binom {n}{r} \frac{\left[1+(-1)^{r+1}\right]}{(r+1)(r+3)}\,|k-k'|^{r+1}(k+k')^{n+1-r}\left[k^2+k'^2+2k\,k' (2+r)\right]\nonumber \\ 
&\quad =\frac{1}{2^{n+2}(n+1)(n+3)(n+4)} \left\{2^{n+3} \left[k^{n+3}
   (k+k' (n+4))+k'^{n+3} (k(n+4)+k')\right] - (n+5)(k+k')^{n+4}\right\}.
\end{align}
\end{subequations}
With the help of the binomial expansion it is now straightforward to show that
\be
\label{eq:ident}
   (k+k')^{n+4}-k^{n+4}-k^{n+3}k' (n+4)-k'^{n+4} -k'^{\,n+3}k(n+4) 
   = k^2 k'^2 \sum_{m=0}^n \binom {n+4}{m+2} k^{n-m}k'^m\,.
\ee
We finally obtain
\be
\begin{split}
   \mathcal{C}_\mathrm{gain}^{(n)}(t) &= \frac{2\st}{a^{6+n}(t)}\,\frac{1}{(n{+}1)(n{+}3)(n{+}4)}\,
   \int_0^\infty \frac{dk}{2\pi^2}\int_0^\infty \frac{dk'}{2\pi^2}f_{k} f_{k'}\, 
   k^2 k'^2 \sum_{m=0}^n \binom {n{+}4}{m{+}2} k^{n-m}k'^m \\
&= 2 \sigma_T \sum_{m=0}^n \frac{(n{+}2) \,n!}{(m{+}2)! (n{-}m{+}2)!} \,\rho_{n-m}(t)\, \rho_m(t)\,.
\end{split}
\ee
This expression determines the term $\mathcal{C}_\mathrm{gain}^{(n)}(t)$~%
\eqref{eq:gain2} in terms of the moments of the distribution function.


\section{Some properties of the associated Laguerre polynomials}

\label{app:Lague} 

In this appendix we collect some useful properties of the associated
Laguerre polynomials that were used in the main text. A broader discussion
of the Laguerre polynomials can be found in Ref.~\cite{Arfken}. \newline

\noindent The closed form of the associated Laguerre polynomial of degree $n$
is 
\begin{equation}  \label{eq:lagpoly}
\mathcal{L}_n^{(\beta)}(x)=\sum_{i=0}^n\,(-1)^i\binom{n+\beta}{n-i}\,\frac{%
x^i}{i!}\,.
\end{equation}
These polynomials satisfy the following orthogonality property 
\begin{equation}  \label{eq:lag-1}
\int_0^\infty\,dx\,e^{-x}\,x^\beta\,\mathcal{L}_n^{(\beta)}(x)\,\mathcal{L}%
_m^{(\beta)}(x) =\frac{(n+\beta)!}{n!}\delta_{nm}\,.
\end{equation}
The generating function of the Laguerre polynomials can be written as 
\begin{equation}  \label{eq:lag-2}
\mathcal{G}(z,x,\beta)\,\equiv\,\sum_{n=0}^\infty\,z^n\,\mathcal{L}%
_n^{(\beta)}(x)\,=\,\frac{e^{-xz/(1-z)}}{(1-z)^{\beta+1}}\,,
\end{equation}
which is valid for $|z|<1$. From the previous expression it is
straightforward to show the following identity 
\begin{equation}  \label{eq:lag-3}
\mathcal{G}(z,x,\beta)-z\frac{\partial \mathcal{G}(z,x,\beta)}{\partial z} =
\sum_{n=0}^\infty\,(1-n)\,z^n\,\mathcal{L}_n^{(\beta)}(x) = \frac{e^{-xz/(1-z)}}{%
(1-z)^{\beta+3}}\left[1+z\left(x+(2+\beta)z-3-\beta\right)\, \right]\,.
\end{equation}


\section{Derivation of Eq.~\eqref{eq:cnmomeq}}

\label{app:nonlincn} 

In this appendix we present the details of the derivation of the Laguerre
moments $c_n$~\eqref{eq:cnmomeq}. From their definition~\eqref{eq:Lagcoeff}
we can find their evolution by taking the derivative respect to $\tau$ 
\begin{equation}
\sum_{r=0}^{n} \,(-1)^r\,\binom {n}{r}\,\partial_\tau M_r =\sum_{r=0}^{n} \,(-1)^r\,%
\binom {n}{r} \left[-M_r +\frac{1}{r{+}1}\sum_{m=0}^{r} M_m(\tau\,) M_{r-m}%
\right],
\end{equation}
where we used explicitly the equation for the normalized moments~%
\eqref{eq:KWmom}. The last expression can be rewritten as 
\begin{equation}  \label{eq:lhs-1}
\!\!\!\!\!\! \partial_\tau c_n + c_n=\sum_{r=0}^{n} \,(-1)^r\,\binom {n}{r}\, \left[%
\frac{1}{r{+}1}\sum_{m=0}^{r}\, M_m\, M_{r-m} \right].
\end{equation}

Now all that remains to be proven is that the RHS of the previous expression
corresponds exactly to the RHS of the Laguerre moment equation~%
\eqref{eq:cnmomeq}. In order to see the equivalence we substitute the
definition of the Laguerre moments~\eqref{eq:Lagcoeff} on the RHS of Eq.~%
\eqref{eq:cnmomeq}: %
\begin{equation}  \label{eq:lhs-2}
\begin{split}
\frac{1}{n+1}\sum_{r=0}^{n} c_r \,c_{n-r}&=\frac{1}{n+1}\,\sum_{r=0}^{n} %
\left[\sum_{s=0}^r \,(-1)^s\,\binom {r}{s}\,M_s\right]\, \left[\sum_{t=0}^{n{%
-}r} \,(-1)^t\,\binom {n{-}r}{t}\,M_t\right]\,, \\
&=\frac{1}{n+1}\,\sum_{s=0}^n\,\sum_{t=0}^n (-1)^{t+s}\,M_s\,M_t\, \left[%
\sum_{r=0}^n \,\binom {r}{s}\,\binom{n{-}r}{t}\right]\,, \\
&=\,\sum_{s=0}^n\,\sum_{t=0}^n (-1)^{t+s}\frac{n!}{(s{+}t{+}1)!\,(n{-}s{-}t)!%
}\,M_s\,M_t =\sum_{q=0}^n\,(-1)^q\,\binom {n}{q}\, \left[\frac{1}{q+1}%
\sum_{s=0}^q\,\,M_s\,M_{q-s}\right].
\end{split}%
\end{equation}
%
This shows that the RHS of Eqs.~\eqref{eq:lhs-1} and~\eqref{eq:lhs-2} agree,
and thus that the equation for the Laguerre moments~\eqref{eq:cnmomeq} holds
for any value of $n\geq 2$. In Eq.~\eqref{eq:lhs-2} we used the fact that $%
\binom {n}{k}=0$ if $k>n$ and the combinatorial identity 
\begin{equation}
\sum_{r=0}^n \,\binom {r}{s}\,\binom{n-r}{t}=\binom{n+1}{s+t+1}\,
\end{equation}
which is valid if $s,t\geq 0$ and $s+t\leq n$. 
\vspace*{-3mm}
\end{widetext}


\bibliography{nonlinear_boltzmann}

\end{document}